\documentclass[twocolumn,apj]{openjournal}

\usepackage[breaklinks,colorlinks,citecolor=blue,urlcolor=blue]{hyperref}
\usepackage{amsmath}

\usepackage[T1]{fontenc}
\usepackage{newtxtext,newtxmath}
\usepackage{graphicx}
\usepackage{amsfonts}
\usepackage{soul}
\usepackage{enumitem}
\usepackage[hyphens]{xurl}
\usepackage{multirow}
\newcommand{\Romannum}[1]{\MakeUppercase{\romannumeral #1}}

\newcommand{\orcidauthor}[3]{\author{\href{http://orcid.org/#1}{#2$^{#3}$}}}

\def\msun{\hbox{${\rm M}_{\odot}$}}
\def\mstar{\hbox{${M}_{\star}$}}

\shorttitle{A Population of Post-Starburst UDGs}
\shortauthors{Sandoval Ascencio et al.}

\begin{document}


\title{\vspace{-0.1cm}Caught in the Act of Quenching? -- A Population of Post-Starburst Ultra-Diffuse Galaxies$^{*}$\vspace{-1.5cm}}

\orcidauthor{0000-0001-8568-8729}{Loraine Sandoval Ascencio}{1,^\dagger} 
\orcidauthor{0000-0003-1371-6019}{M. C. Cooper}{1}
\orcidauthor{0000-0002-5177-727X}{Dennis Zaritsky}{2}
\orcidauthor{0000-0001-7618-8212}{Richard Donnerstein}{2} 
\orcidauthor{0000-0002-7013-4392}{Donghyeon J. Khim}{2} 
\orcidauthor{0000-0002-8209-2783}{Devontae C. Baxter}{3,^\ddagger}

\affiliation{$^{1}${Department of Physics \& Astronomy, University of California, Irvine, 4129 Reines Hall, Irvine, CA 92697, USA}} 
\affiliation{$^{2}${Steward Observatory and Department of Astronomy, University of Arizona, 933 N. Cherry Avenue, Tucson, AZ 85721, USA}} 
\affiliation{$^{3}${Department of Astronomy \& Astrophysics,
University of California, San Diego, 9500 Gilman Dr, La Jolla, CA 92093, USA}}

\thanks{$^\dagger$ Corresponding author: \href{mailto:lorainas@uci.edu}{lorainas@uci.edu}}
\thanks{$^\ddagger$ NSF Astronomy and Astrophysics Postdoctoral Fellow}
\thanks{$^*$ The data presented herein were obtained at the W. M. Keck Observatory, which is operated as a scientific partnership among the California Institute of Technology, the University of California and the National Aeronautics and Space Administration. The Observatory was made possible by the generous financial support of the W. M. Keck Foundation.}

\begin{abstract} 
    We report the discovery of post-starburst ultra-diffuse galaxies (UDGs), identified through spectroscopic analysis with KCWI at the Keck \Romannum{2} Telescope. 
    Our analysis is based on a sample of 44 candidate UDGs selected from the Systematically Measuring Ultra-Diffuse Galaxies (SMUDGes) program. 
    Our measured spectroscopic redshifts reveal $\sim 85\%$ of the entire KCWI sample exhibit large physical sizes ($R_{e} \gtrsim 1~{\rm kpc}$) and low surface brightnesses ($24 \lesssim \mu_{0,g} \lesssim 25$ mag arcsec$^{-2}$) which categorize them as UDGs. 
    We find $20\%$ of the confirmed UDG population contain post-starburst (or K+A) features, characterized by minimal to no emission in H$\beta$ indicative of quenched star formation and a predominant presence of spectral A-type stars. 
    In surveying the local environments of the post-starburst UDGs, we find that nearly half are isolated systems, which is unusual given that isolated UDGs are most commonly found to be star-forming.
    Two of these systems reside $2-3~R_{\rm vir}$ away from potential nearby massive hosts ($\mstar >10^{10}~\msun$), indicating the absence of environmental influence. 
    These post-starburst UDGs may represent systems experiencing star formation feedback such that a recent burst may lead to (at least temporary) quenching.
    Overall, our results highlight the potentially diverse quenching pathways of UDGs in the local Universe.

    \keywords{galaxies:dwarf -- galaxies:star formation -- galaxies:evolution -- galaxies:structure}
\end{abstract}

\section{Introduction}
A decade ago, the identification of a large population of ultra-diffuse galaxies (or UDGs) within the Coma Cluster \citep{vanDokkum2015a, vanDokkum2015b} sparked a resurgence of interest in the study of low surface brightness galaxies \citep[e.g.][]{Sandage1984, Bothun1991, deBlok1997, Impey1997}.
Ultra-diffuse galaxies are unique in that they have stellar masses akin to that of dwarf galaxies ($\mstar \sim 10^{7.5-8.5}~\msun$), but the effective radii of much larger systems ($R_{e} \gtrsim 1-5~{\rm kpc}$, comparable to that of the Milky Way). These extreme galaxies, which have typical surface brightnesses fainter than $24$~mag~${\rm arcsec}^{-2}$, challenge our understanding of galaxy formation and have accordingly been the subject of great interest over the last decade.
Subsequent searches in other nearby clusters (other than Coma) have found numerous UDGs in cluster environments, with the population largely having no active star formation (i.e.~quenched, \citealt{Mihos2015, Koda2015, vanderBurg2016, vanderBurg2017, Yagi2016}).

Various formation scenarios have been proposed, including the idea that these systems are somehow failed Milky Way-like galaxies that reside in massive 
dark matter halos but fall short of forming most of their expected stellar content due to feedback \citep{vanDokkum2016}. 
Meanwhile, other potential formation mechanisms posit that UDGs form as a result of galaxy interactions \citep{Bennet2018} or within halos in the tail of the spin distribution \citep{Amorisco2016, Leisman2017, Benavides2023} or as a consequence of extremely strong star formation feedback in dwarf galaxies \citep{Chan2018, DiCintio2017, Freundlich2020}. 

Several recent studies have illustrated that tidal heating of dwarf galaxies within massive clusters can effectively puff-up systems, reproducing the observed sizes, stellar masses, and number density of UDGs within nearby groups and clusters \citep{Carleton2019, Sales2020, Tremmel2020, Jones2021, Fielder2024}. 
One uncertainty in modeling the formation of UDGs is the connection between quenching and size evolution, including whether UDGs are quenched prior to infall and subsequent tidal heating. 
Moreover, while tidal heating is successful in reproducing many properties of the UDG populations in (or near) massive halos \citep{Benavides2021}, it is unlikely to explain the origins of isolated (field) UDGs.

While the vast majority of UDG searches have focused on cluster populations \citep[e.g.][]{Gannon2022, Bautista2023, Forbes2023, Carleton2023}, some work has targeted UDGs in lower-density environments 
\citep[e.g.][]{Roman2017a, Roman2017b, Shi2017, Gannon2021, Marleau2021}, with very few isolated UDGs identified.  
All known isolated UDGs (or those in low-density environments) are star-forming systems, in contrast to their counterparts in groups and clusters \citep{Leisman2017, Rong2020a}.
Passive, isolated  UDGs would represent a particular challenge to current formation models. Not only is tidal stripping ineffective in isolation, stellar feedback struggles to directly produce quenched UDGs \citep[e.g.][]{Chan2018, Wright2021}.

\begin{figure*}
	\centering
    \includegraphics[width=\linewidth]{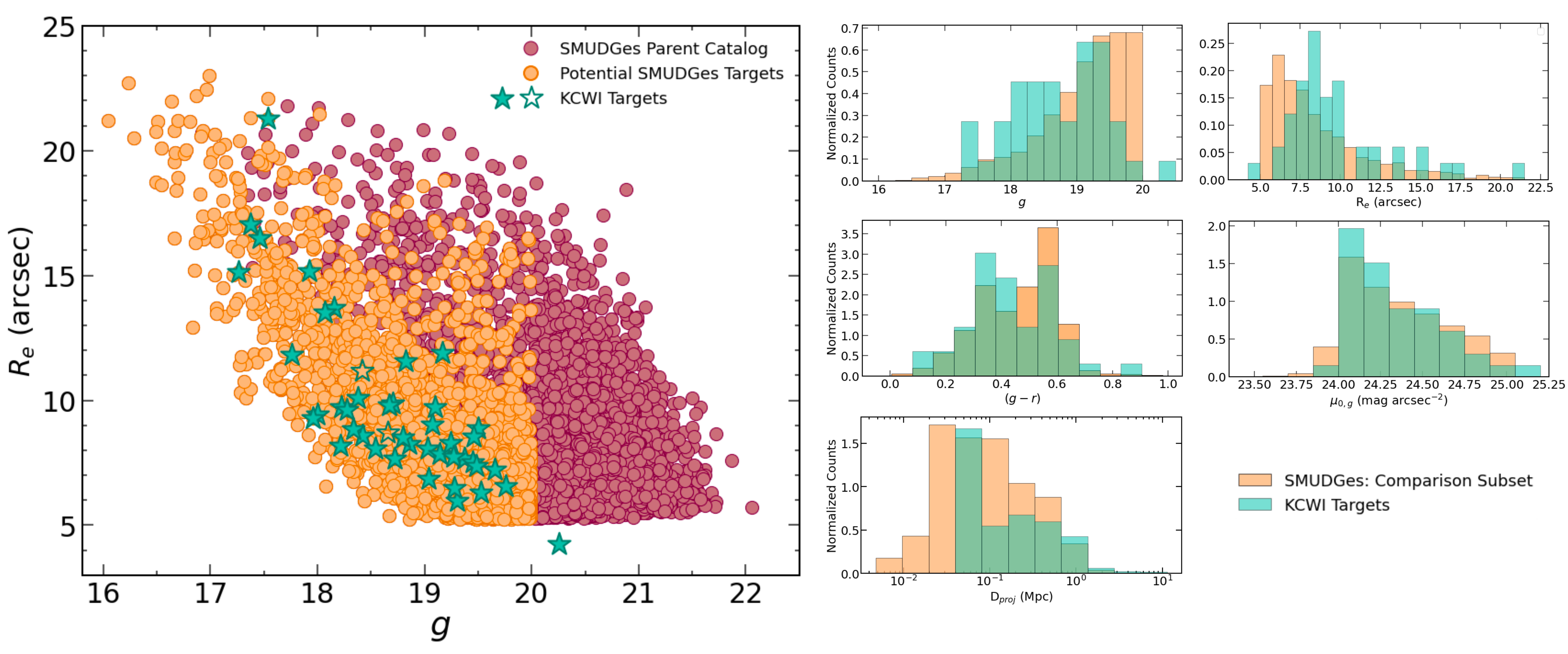}
		\caption{(\emph{Left}) The distribution of size versus apparent $g$-band magnitude for our KCWI targets (green closed and open stars) in comparison to the SMUDGes parent catalog (red and orange circles). The KCWI targets were selected from a subset of the SMUDGes catalog (orange circles), defined by $g < 20$, $R_{e} > 5^{\prime\prime}$, and $\mu_{0,g}<25$ mag arcsec$^{-2}$. The two KCWI targets plotted as open stars are the two systems without a secure spectroscopic redshift. (\emph{Right}) The distribution of apparent $g$-band magnitude, angular size, apparent $g-r$ color, central surface brightness, and projected distance from nearest massive ($>10^{10}~\msun$) galaxy for our KCWI targets relative to the subset of the SMUDGes catalog meeting our target selection criteria.}
\label{fig:targetcomp}
\end{figure*}

Finding recently-quenched UDGs may offer a unique opportunity to probe the environments and potential mechanisms by which star formation is suppressed within this extreme galaxy population. As such, the discovery of a population of post-starburst (otherwise known as E+A or K+A galaxies) could be impactful. 
Post-starburst galaxies are commonly identified by their lack of ongoing star formation, combined with the presence of a young A-type stellar population that is indicative of recent star formation \citep[e.g.][]{Dressler1983, Yang2004}. 
Thus, post-starburst galaxies are selected to be systems where quenching has very recently occurred, offering a unique probe into the relevant mechanisms at play in suppressing star formation and the evolutionary history of the galaxy population. 

Herein, we present results from a spectroscopic follow-up of UDG candidates selected from wide-field imaging surveys, aiming to probe a range of environments and levels of star-formation activity. 
In \S\ref{sec:data}, we describe our target sample, observations, and data reduction along with spectral fitting performed to identify potential post-starburst sources. 
Meanwhile, in \S\ref{sec:results}, we examine the classification of our targeted sources as UDGs and present the discovery of a population of post-starburst UDGs. 
In \S\ref{sec:discussion}, we analyze the local environments of the post-starburst UDG population and discuss potential quenching mechanisms. 
%
Finally, we summarize our primary results in \S\ref{sec:summary}. 
Throughout this work, we adopt a flat $\Lambda$CDM cosmology with $\Omega_{m}=0.3$ and H$_{0}=74$~km~s$^{-1}$~Mpc$^{-1}$. 
All magnitudes, unless otherwise stated, are given on the AB system \citep{Oke1983}.

\begin{figure*}
\centering
\includegraphics[width=1\textwidth]{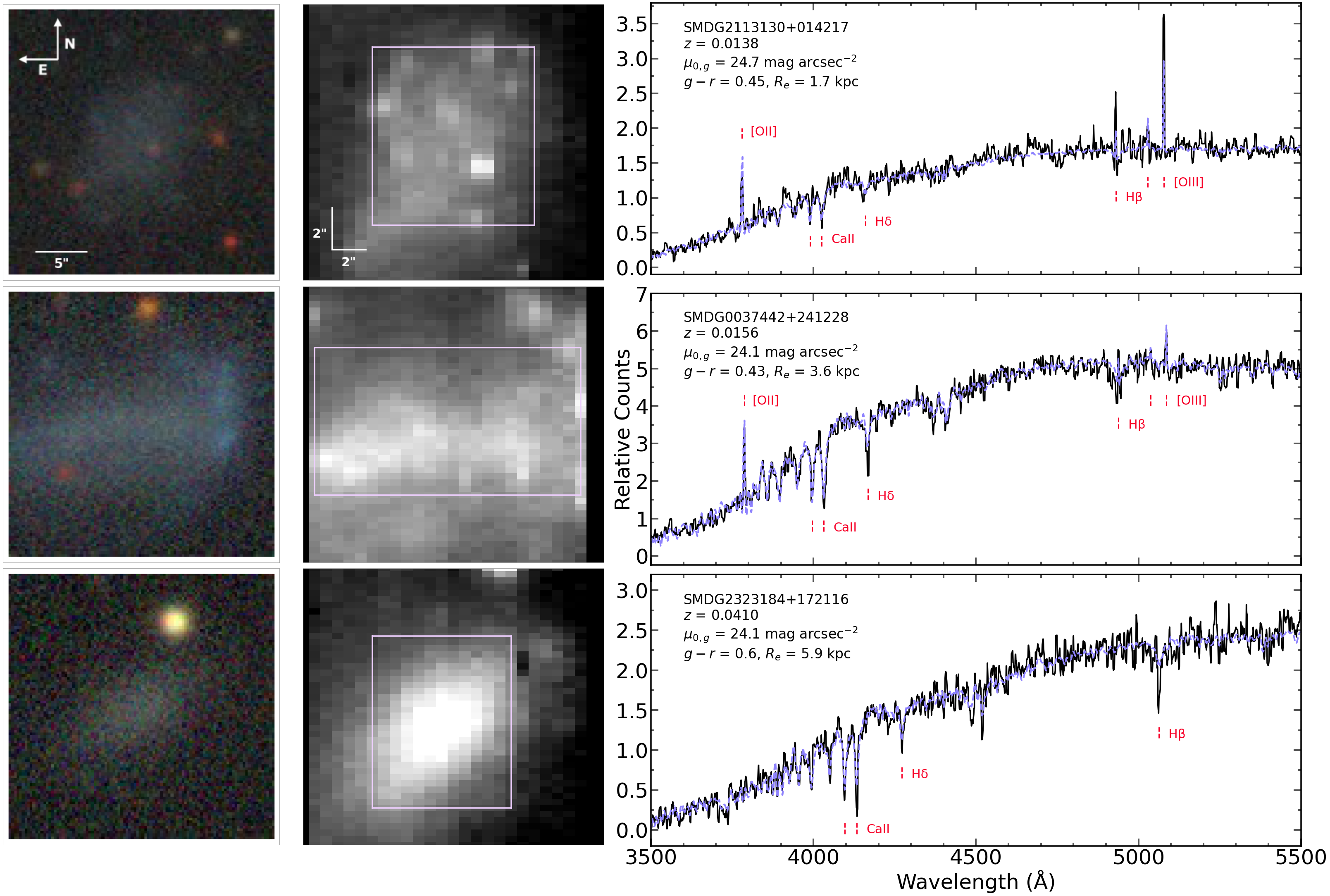}
\caption{For three representative targets from our KCWI sample, we show Legacy Sky Survey $gri$ images (\emph{left}), KCWI 2-d collapsed spectral images (\emph{center}), and extracted KCWI 1-d spectra (\emph{right}). The purple rectangles in the center panel show the window within which the 1-d spectra are extracted. The 1-d spectra (black lines) are smoothed using an inverse-variance weighting and a kernel of 5~\AA{}. For comparison, the best-fit spectra from \texttt{Redmonster} is included as the dashed purple line, and the location of prominent spectral features are indicated by the red dashed vertical lines.
}
\label{fig:fittedspectra}
\end{figure*}

\section{Data}
\label{sec:data}

\subsection{Description of Target Sample}
\label{subsec:targets} 
Targets for spectroscopic follow-up were chosen from the Systematically Measuring Ultra-Diffuse Galaxies (SMUDGes) program \citep{zaritsky2019,zaritsky2021,zaritsky2022,zaritsky2023}.
SMUDGes emerged as an effort to search for ultra-diffuse galaxies using multiple imaging campaigns that constitute the Dark Energy Spectroscopic Instrument (DESI) Legacy Imaging Surveys \citep{dey2019}, hereafter referred to as the Legacy Survey. 
The SMUDGes catalog contains nearly $7000$ UDG candidates spanning nearly $20{,}000$ square degrees of sky, utilizing imaging from the CTIO/DECam (via DECaLs, \citealt{Flaugher2015}), the Mayall $z$-band Legacy Survey \citep[MzLS,][]{Dey2016}, and the Beijing-Arizona Sky Survey \citep[BASS,][]{Zhou2017}. 
Sources within the SMUDGes catalog are identified as \emph{candidate} UDGs based on their measured \emph{angular} size. Follow-up spectroscopy with KCWI is intended to provide a spectroscopic redshift and thus a robust measurement of \emph{physical} size.

Our targets were selected from SMUDGes to be bright in the optical ($g < 20$), have an effective radius of $R_{e}>5^{\prime\prime}$, and have a $g$-band central surface brightness of $\mu_{0,g}$ < 25 mag arcsec$^{-2}$.
Figure~\ref{fig:targetcomp} (\textit{left}) illustrates the distribution of angular sizes and apparent magnitudes for our sample relative to the parent SMUDGes catalog and the sample of all SMUDGes sources meeting the target selection criteria. 
Note that one targeted source (SMDG0134336+005425) has an apparent size and magnitude outside of our selection limits, as its selection was based on an earlier version of the SMUDGes catalog. 
As shown in Figure~\ref{fig:targetcomp} (\emph{right}), within the limits of the target selection criteria, our sample is biased slightly towards some of the brighter and larger SMUDGes sources. 
Our targets were also selected to avoid proximity to bright objects that could potentially impact the ability to perform local sky subtraction. 
Altogether, our spectroscopic sample is representative of the broader SMUDGes data set (within the selection limits), spanning a broad range of apparent color so as to include both star-forming and passive sources.

Finally, Fig.~\ref{fig:targetcomp} also compares the distribution of local environment for the KCWI targets relative to that of the larger population of potential SMUDGes targets. As a measure of local environment, we utilize the projected distance ($D_{\rm proj}$) to the nearest massive galaxy, with our local massive galaxy sample drawn from v1\_0\_1 of the NASA-Sloan Atlas \citep[NSA,][]{Blanton2011, Geha2012}.\footnote{\href{http://www.nsatlas.org}{http://www.nsatlas.org}}
We select massive galaxies with $\mstar > 10^{10}~\msun$ and limit the sample to systems at $z < 0.06$, to overlap with the most probable redshift range of SMUDGes sources. Our results are qualitatively unchanged when relaxing this redshift limit (e.g. $z < 0.1$). 
Given that the NSA does not cover the same footprint as SMUDGes, we limit the comparison sample to a smaller area ($\sim 5000$~square~degrees) where the NSA provides a robust tracer population. This roughly halves the SMUDGes comparison sample and excludes $4$ KCWI targets that reside near or beyond the edge of the NSA footprint. 
Overall, the KCWI targets span a range of environments, including near to and far away from potential hosts.
However, our KCWI targets were intentionally excluded from regions near the Coma and Virgo clusters, which partially accounts for the relative bias in the distribution of local environment measures for the KCWI targets and the SMUDGes parent sample as shown in Fig.~\ref{fig:targetcomp}.
Roughly $38\%$ of the SMUDGes comparison sample (within the $\sim 5000$~deg$^2$ footprint) is located within $\sim 2$~Mpc (projected) of these well-known, nearby clusters.

\subsection{KCWI Observations and Data Reduction}
\label{subsec:kcwi}
Our spectroscopic follow-up effort included integral field spectroscopy of 44 UDGs using the Keck Cosmic Web Imager \citep[KCWI,][]{Martin2010, Morrissey2018} on Keck \Romannum{2}, with observations collected in the fall of 2020 and 2021. We utilized the medium image slicer with the BL grating centered at 4500~\AA{} and a peak wavelength of 4200~\AA{} for a field-of-view of roughly $16\arcsec{\times}20\arcsec$ and spectral resolution of R$\sim1800$ over the wavelength range of $3500$ to $5500$~\AA{}.
The total exposure time for each source varied from $1200$~{s} to $4500$~{s}, spanning a minimum of two science frames (see Table~\ref{tab:kcwi} for more detail). 
In addition, for each source, we acquired a minimum of one sky frame, offset by $\sim 60^{\prime\prime}$ and with an exposure time typically matched to that of an individual science frame, for use in non-local sky subtraction. 
Each night, standard calibrations were acquired, including bias, arc, and flat frames.

Our data were reduced using the KCWI data reduction pipeline (KCWI DRP\footnote{\url{https://kcwi-drp.readthedocs.io/en/latest/}}). For most of our targets, we utilized non-local sky subtraction by specifying a sky frame. For a subset ($11$) of our targets, however, a higher signal-to-noise spectrum was achieved via local sky subtraction. 
From the reduced data cubes, we extracted a 1-d spectrum for each target, with the spatial extent of the extraction window selected to maximize the signal-to-noise of the resulting spectrum (as measured at $4500$-$4750$~\AA) and weighting by the inverse-variance during co-addition. 
Figure~\ref{fig:fittedspectra} shows 2-d images from the Legacy Survey alongside the collapsed co-added data cube images and the corresponding extracted 1-d spectra for three of our sources.

\subsection{Measurement of Spectroscopic Redshifts, Stellar Masses, and Physical Sizes}
\label{subsec:redshift}
To measure spectroscopic redshifts, we use the redshift measurement and spectral classification software \texttt{Redmonster}\footnote{\url{https://github.com/timahutchinson/redmonster}} \citep{Hutchinson2016} to fit theoretical galaxy spectral templates to the KCWI spectrum of each source by means of a $\chi^{2}$-minimization procedure. 
For each source, the top five best-fit prospective spectroscopic redshifts were visually inspected, yielding a secure redshift for $42$ out of $44$ targets. 
We consider redshifts to be secure for sources where the H and K calcium absorption lines are clearly identified and/or emission lines such as [O{\scriptsize II}], [O{\scriptsize III}], and those from the Balmer series are evident.
Figure~\ref{fig:fittedspectra} shows the reduced 1-d spectra for 3 representative KCWI targets, along with the best-fit \texttt{Redmonster} template.

Using the SMUDGes photometry and the measured heliocentric-corrected redshift (and corresponding luminosity distance), we estimate absolute magnitudes in the $g$ and $r$ bands for each KCWI target with a secure spectroscopic redshift, taking into account Galactic extinction \citep{Schlegel1998, Green2018}. 
Following the mass-to-light versus color relation from \citet{Roediger2015}, we then utilize the extinction-corrected $g-r$ color to calculate the $\mstar/L_{r}$ for each source and scale by the absolute $r$-band magnitude to estimate the stellar mass. 
Using {\sc PROSPECTOR} \citep{Johnson2021}, we performed SED fitting to derive stellar masses for the entire SMUDGes catalog. Results and details on the procedure will be presented in Zaritsky et al. (in prep). We incorporated photometric data from DECaLS in the \textit{grz} bands \citep{Flaugher2015}, GALEX at far- and near-ultraviolet wavelengths \citep[FUV and NUV,][]{GildePaz2007}, and WISE in the $W_1$, $W_2$, $W_3$, and $W_4$ bands \citep{Wright2010}. 
We performed this SED fitting wherever sufficient photometric coverage was available, which was the case for 29/44 KCWI sources.
Adopting a non-parametric star formation history (SFH), we find that stellar masses calculated using {\sc PROSPECTOR} are in good agreement with those based on the mass-to-light versus color relation, with a modest RMS difference of $\sim 0.4$~dex, such that the SED fitting favors slightly larger stellar masses. 
Given the relatively small offset and the simplicity of the color-based approach, we adopt stellar masses derived from the mass-to-light versus color relation for our analysis.
We employ our estimated spectroscopic redshifts to convert the measured angular sizes ($R_{e}$) -- as measured from stacked images using a combination of $grz$ bands -- to physical sizes measured in kpc.
The KCWI targets, along with their corresponding heliocentric-corrected redshifts and other properties, are detailed in Table~\ref{tab:kcwi}.

\begin{table}
	\setlength{\tabcolsep}{3pt}
	\centering
    \caption{Properties of KCWI Targets}
    \begin{tabular}{c c c c c c}
	    \hline \hline
	    \multirow{2}{*}{Source ID\textsuperscript{a}} & Observation & $t_{\rm exp}$ & \multirow{2}{*}{$z_\text{helio}$\textsuperscript{d}} & \multirow{2}{*}{$\log(\frac{\mstar}{\msun})$\textsuperscript{e}} & $R_{e}$  \\
	     & Date\textsuperscript{b} & (s)\textsuperscript{c} &  &   & (kpc)\textsuperscript{f} \\
	    \hline
SMDG0000334+165424  &  2021-09-15  &  2700  &  ---  &  ---  &  --- \\
SMDG0015089-031837  &  2021-10-04  &  1200  &  0.02078  &  8.2  &  4.7 \\
SMDG0015208+200027  &  2021-10-30  &  2400  &  0.00768  &  7.6  &  1.4 \\
SMDG0033164+283311  &  2021-10-04  &  1800  &  0.0163  &  7.8  &  2.6 \\
SMDG0037442+241228  &  2021-10-04  &  1200  &  0.01559  &  8.5  &  3.6 \\
SMDG0038365-064207  &  2021-10-03  &  1800  &  0.01067  &  8.6  &  3.5 \\
SMDG0045533-093353  &  2021-10-03  &  1800  &  0.01961  &  8.8  &  3.0 \\
SMDG0102539+305357  &  2021-10-04  &  1200  &  0.01383  &  8.1  &  2.2 \\
SMDG0109254-014543  &  2020-09-15  &  4500  &  0.01349  &  8.9  &  3.0 \\
SMDG0122562+084025  &  2021-10-04  &  3000  &  0.00822  &  7.8  &  1.4 \\
SMDG0124406-013812  &  2020-09-15  &  3600  &  0.01695  &  8.4  &  2.5 \\
SMDG0127038-082400  &  2021-10-03  &  1200  &  0.00663  &  7.3  &  1.1 \\
SMDG0134336+005425  &  2020-09-15  &  2940  &  0.04403  &  7.8  &  3.5 \\
SMDG0136469+034323  &  2021-10-04  &  3000  &  0.00552  &  7.2  &  1.5 \\
SMDG0145431-084143  &  2021-10-03  &  1200  &  0.00638  &  7.5  &  1.2 \\
SMDG0147005-142556  &  2021-10-03  &  3000  &  0.00587  &  7.2  &  1.0 \\
SMDG0148234-131444  &  2021-10-03  &  2700  &  0.00573  &  7.2  &  1.0 \\
SMDG0152163-044242  &  2021-09-15  &  1200  &  0.01693  &  8.4  &  6.9 \\
SMDG0224231-031059  &  2021-10-03  &  2400  &  0.00462  &  7.6  &  1.4 \\
SMDG0321028-042741  &  2021-10-04  &  1250  &  0.01369  &  8.1  &  2.3 \\
SMDG0455129-031122  &  2021-09-15  &  2160  &  0.01437  &  8.1  &  2.1 \\
SMDG0459164-072243  &  2021-10-04  &  3600  &  0.01331  &  8.2  &  2.5 \\
SMDG1756253+331055  &  2021-10-04  &  1800  &  0.01633  &  8.2  &  4.3 \\
SMDG2104498+002509  &  2020-09-15  &  2400  &  0.01350  &  6.9  &  1.9 \\
SMDG2113130+014217  &  2020-09-15  &  2400  &  0.01378  &  7.7  &  1.7 \\
SMDG2207591-100713  &  2021-10-03  &  1800  &  0.00995  &  7.5  &  1.6 \\
SMDG2217328-132522  &  2021-10-03  &  1800  &  0.00554  &  7.4  &  1.7 \\
SMDG2227407-122104  &  2021-10-03  &  1200  &  0.01601  &  8.1  &  3.1 \\
SMDG2235086+011040  &  2020-09-15  &  2700  &  0.05631  &  9.2  &  6.8 \\
SMDG2235405+233627  &  2021-10-04  &  1200  &  0.00411  &  7.3  &  0.8 \\
SMDG2239446+180218  &  2021-10-04  &  1200  &  0.02343  &  8.4  &  4.4 \\
SMDG2240062+223715  &  2021-10-30  &  4800  &  0.0043  &  6.9  &  0.6 \\
SMDG2241452+222153  &  2021-10-04  &  1200  &  0.02567  &  8.3  &  3.8 \\
SMDG2246485-105427  &  2021-10-03  &  2400  &  ---  &  ---  &  --- \\
SMDG2251037-015247  &  2020-09-15  &  4350  &  0.00996  &  7.8  &  1.2 \\
SMDG2300066+154949  &  2021-10-30  &  1800  &  0.00708  &  7.4  &  0.9 \\
SMDG2302155+300423  &  2021-10-30  &  1800  &  0.00278  &  6.0  &  0.4 \\
SMDG2323184+172116  &  2021-10-30  &  1800  &  0.04103  &  9.2  &  5.9 \\
SMDG2329214-063209  &  2021-10-04  &  1200  &  0.03223  &  8.8  &  5.0 \\
SMDG2337078+001240  &  2021-10-04  &  1800  &  0.00857  &  8.0  &  1.7 \\
SMDG2338066+003203  &  2021-10-04  &  1200  &  0.03314  &  8.7  &  5.3 \\
SMDG2343135-083901  &  2021-10-03  &  2700  &  0.00627  &  7.5  &  1.2 \\
SMDG2359136+164655  &  2021-10-30  &  1800  &  0.00344  &  7.1  &  0.5 \\
SMDG2359282+150125  &  2021-09-15  &  1800  &  0.00581  &  7.6  &  1.9 \\
 
 \hline
\multicolumn{6}{l}{\textsuperscript{a}\footnotesize{SMUDGes ID \citep{zaritsky2023}}} \\ 
\multicolumn{6}{l}{\textsuperscript{b}\footnotesize{UT Date of the KCWI observation}} \\
\multicolumn{6}{l}{\textsuperscript{c}\footnotesize{total KCWI exposure time (seconds)}} \\
\multicolumn{6}{l}{\textsuperscript{d}\footnotesize{heliocentric-corrected redshift}} \\
\multicolumn{6}{l}{\textsuperscript{e}\footnotesize{stellar mass}} \\
\multicolumn{6}{l}{\textsuperscript{f}\footnotesize{effective radius in physical kpc }} \\
\end{tabular}
\label{tab:kcwi}
\end{table}

\begin{figure*}
\centering
\includegraphics[width=0.7\textwidth]{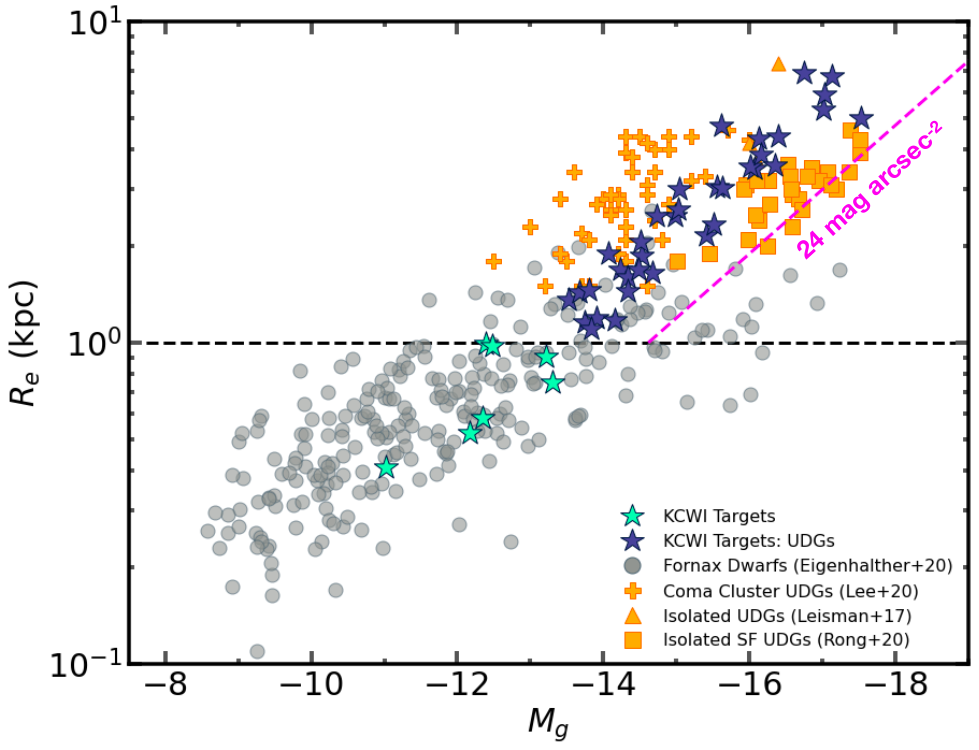}
\caption{Relationship between half-light radius and $g$-band absolute magnitude for our KCWI sample (green and blue stars) relative to existing UDG (orange symbols, \citealt{vanDokkum2015a, Rong2020a, Leisman2017}) and dwarf (grey circles, \citealt{Eigenthaler2018}) populations in the local Universe. The black dashed line at $R_{e} = 1$~kpc defines our UDG size limit, while the magenta dashed line corresponds to a surface brightness limit of $\mu_{g}=24$~mag~arcsec$^{-2}$.  }
\label{fig:sizemag}
\end{figure*}

\subsection{Spectral Fitting}
\label{subsec:specfit}
Various approaches have proven effective in identifying post-starburst features.
For example, some efforts have applied tools such as {\sc BAGPIPES} \citep{Carnall2018} and {\sc PROSPECTOR} to model galaxy stellar populations and star formation histories, leveraging both spectroscopic and photometric data \citep[e.g.][]{Wild2020, Bezanson2022, Suess2022}. 
Meanwhile, other studies utilize principal component analysis as applied to multi-wavelength photometry \citep{Wild2014} or machine learning \citep[e.g.][]{Meusinger2017, Baron2017}.
Finally, as demonstrated by many previous studies in the local Universe \citep[e.g.][]{Dressler1983, Zabludoff1996, Quintero2004}, decomposition of optical spectra -- covering a very similar wavelength regime to that of our KCWI spectra -- has proven extremely effective at identifying galaxies where star formation has recently been quenched (so-called post-starburst, E+A, or K+A galaxies). 
In this work, we adopt the latter approach.
For each KCWI target with a secure spectroscopic redshift, we calculate the relative contribution from A- and K-type stellar populations, by fitting a linear combination of an A- and K-star spectrum to the reduced KCWI 1-d spectrum. 
For our K- and A-star components, we use stellar templates from the Medium resolution Isaac Newton
Library of Empirical Spectra (MILES) stellar library \citep{Sanchez2006} corresponding to stars HD152781 (spectral type K0/K1) and HD027819 (spectral type A7).
We convolve both templates to a velocity width of 75~km~s$^{-1}$, corresponding to the instrumental resolution of KCWI combined with the assumed velocity width of a typical dwarf galaxy.
While we find qualitatively similar results when using alternative templates, most notably a passive galaxy template based on the SDSS Luminous Red Galaxy (LRG) sample from \citet{Eisenstein2003} and a model A0 star (Vega) spectrum from the Kurucz library\footnote{\url{http://kurucz.harvard.edu/stars/vega/veg1000pr25.500000}} (see \citealt{Yoon2010} and references therein), we focus our analysis on spectral fits using the MILES-based templates. 

Before performing our fits, we shift the KCWI spectra to the rest frame and trim the data to a wavelength range of 3700-5500~\AA{}. 
This range is chosen to match the high signal-to-noise region of the spectral templates.
To focus our analysis on the strength of spectral features, such as the Balmer absorption lines, rather than the shape of the continuum emission, we opt to fit and remove the continuum from our KCWI spectra as well as from the A- and K-star templates. 
To define the continuum level, we mask the location of prominent emission and absorption lines -- e.g.~[O{\scriptsize II}] ($\mathrm{\lambda\lambda}$3727, 3729), Ca {\scriptsize II} H and K lines, G-band, Balmer lines, [O{\scriptsize III}] ($\mathrm{\lambda\lambda}$4959, 5007), and Mg {\scriptsize I} b triplet ($\mathrm{\lambda\lambda}$5167, 5173, 5184) -- and then smooth each spectrum using a 250~\AA{}-wide boxcar window.
After continuum subtraction, we perform a least-squares fitting technique to fit a linear combination of the A- and K-star templates to the KCWI spectra, using the coefficients of the fit to quantify the ratio of A- to K-star emission (${\rm A}/{\rm K}$).

As a probe on ongoing star formation in our KCWI targets, we measure the equivalent width (EW) of the H$\beta$ and [O{\scriptsize II}] emission lines. 
To measure the H$\beta$ emission, we define a $15$~\AA{} window centered on the line. Following \citet{Quintero2004}, we subtract the best-fit K+A composite template from the KCWI spectrum (so as to account for a potential H$\beta$ absorption trough). Using the continuum-subtracted spectrum, we then fit a single Gaussian to the remaining H$\beta$ emission (or a double Gaussian with a fixed 3:2 line strength ratio for [O{\scriptsize II}]). To measure the line emission, we integrate the resulting Gaussian fit over a $\pm3\sigma$ window (typically $\sim5-10$~\AA). 
For cases where H$\beta$ emission is undetected (or only marginally detected), resulting in a poor Gaussian fit, we limit the extraction window to $12$~\AA{}. 
To compute the emission line EW, we measure the continuum for each KCWI spectrum within two $50$~\AA{} windows on either side of the H$\beta$ line (or [O{\scriptsize II}] doublet),  taking the inverse variance-weighted average in those windows as the continuum level.

Both the ${\rm A}/{\rm K}$ and H$\beta$ EW measurements are relatively insensitive to the adopted extraction window within the KCWI data cube. 
For example, varying the extraction window for each target by $\pm20\%$ in pixel area yields a $1$-$\sigma$ scatter of only $\sim 20\%$ in ${\rm A}/{\rm K}$ ratio and $\sim 25\%$ in H$\beta$ EW, with a few outliers corresponding to systems with pronounced star-forming knots.

\subsection{Comparison SDSS Sample}
\label{subsec:sdss}
As a comparison sample, we select $30{,}000$ low-redshift ($0.1 < z < 0.2$) galaxies from the main galaxy sample of the Sloan Digital Sky Survey DR18 \citep[SDSS,][]{York2000, Almeida2023}. We restrict to this redshift range to ensure that the SDSS spectra cover the same rest-frame wavelength range and spectral features as our KCWI observations (e.g.~[O{\scriptsize II}], Ca H \& K, H$\delta$, H$\beta$, [O{\scriptsize III}]). 
Moreover, this redshift range helps mitigate aperture effects by allowing the SDSS fiber to encompass a larger fraction of the galaxy's light, yielding spectra that are more representative of the galaxy as a whole.
At this redshift range, the comparison sample spans a stellar mass range of $10^{9}-10^{11}~\msun$, based on stellar mass estimates from the MPA-JHU \texttt{galSpecExtra} DR8 catalog for which we crossmatched with the DR18 spectroscopic catalog \citep{Kauffmann2003, Brinchmann2004}. 
In addition, we restrict to sources on plates with ``good'' data quality, based on the SDSS \texttt{PLATEQUALITY} flag.
For this SDSS comparison sample, we performed spectral fitting (${\rm A}/{\rm K}$ and EW measurements) following the same procedures applied to the KCWI spectra (cf.~\S\ref{subsec:specfit}). 
To avoid any potential systematic bias associated with the broader spectral coverage of the SDSS spectra, we limit the spectral range to 3700 - 5500~\AA{}, matching that of the KCWI fitting technique.

\section{Results}
\label{sec:results}

\subsection{Confirmation of UDGs}
\label{subsec:udg}
Typically UDGs are selected according to their large physical size and low surface brightness. 
Various selection criteria, however, have been utilized to identify UDG populations, with size cuts ranging from $R_{e} >0.7~{\rm kpc}$ to $R_{e}>1.5~{\rm kpc}$ \citep[e.g.][]{vanDokkum2015a, Yagi2016, Sales2020} and sometimes with a surface brightness or stellar mass surface density limit applied to exclude more massive galaxies where such large sizes are less atypical \citep[e.g.~$\mu < 24$ mag arcsec$^{-2}$][]{vanderBurg2016, Carleton2019}. 
In Figure~\ref{fig:sizemag}, we show the distribution of physical size (as measured in the $g$-band) versus absolute $g$-band magnitude for our KCWI targets alongside corresponding measurements from several existing UDG and dwarf galaxy samples in the local Universe \citep{vanDokkum2015a, Leisman2017, Eigenthaler2018, Rong2020a}.  
The dashed lines correspond to a size limit of $R_{e} = 1~{\rm kpc}$ and an effective surface brightness limit of $\mu_{g} = 24$ mag arcsec$^{-2}$, which define our KCWI target's UDG selection criteria. 
From the sample of $42$ UDG candidates from SMUDGes, with a secure spectroscopic redshift via our KCWI follow-up spectroscopy, we find that $35$ are classified as UDGs ($\sim 85$\% of our targets). 
Even when adopting a more strict size criterion of $R_{e} = 1.5~{\rm kpc}$, the majority of our targets ($\sim 65$\%) remain classified as UDGs.

\begin{figure*}
\centering
\includegraphics[width=0.85\textwidth]{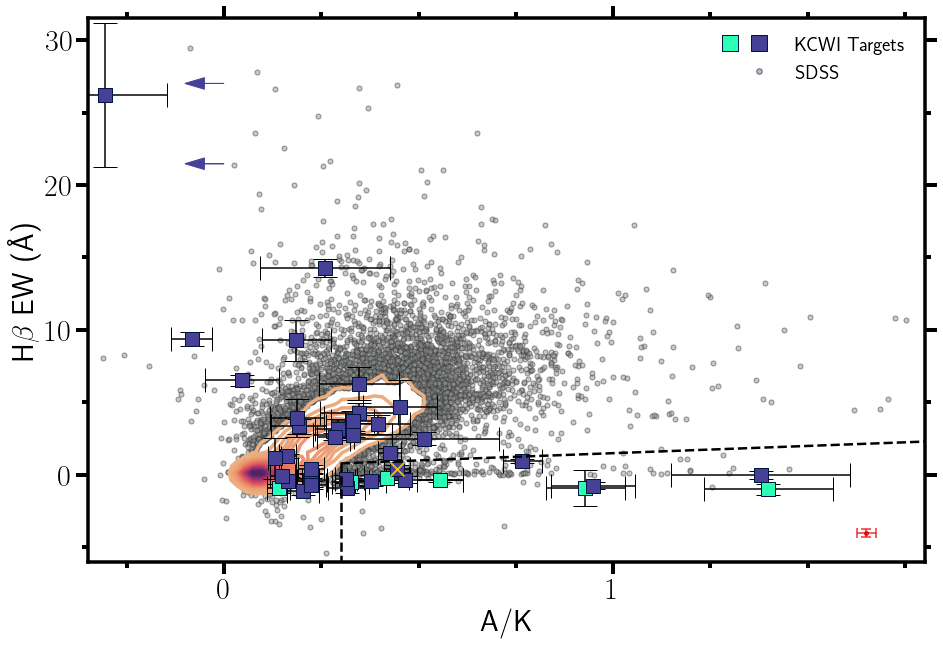}
\caption{The distribution of H$\beta$ EW versus ${\rm A}/{\rm K}$ ratio for our KCWI sample (shaded squares) relative to a sample of $30{,}000$ galaxies at $0.1 < z < 0.2$ selected from the SDSS (contours and grey circles). 
The dashed lines illustrate our post-starburst selection, which identifies $13$ post-starburst candidates including $8$ post-starburst UDGs. Sources that meet the UDG criteria of $R_{e} > 1$~kpc are colored in blue, while those below the UDG size threshold are shaded green.  
The source that meets the post-starburst selection -- but has significant [O{\scriptsize II}] emission -- is highlighted using a gold cross. In the lower right corner (in red), we show the median errors for the SDSS sample.
Prior to EW measurements, underlying stellar absorption was subtracted from all spectra.
}
\label{fig:EWAK}
\end{figure*}

\subsection{Discovery of Post-Starburst UDG Population}

The hallmark of post-starburst galaxies is a lack of ongoing star formation combined with the presence of a population of young (short-lived) stars indicating recent star formation. 
In most searches for post-starbursts at low redshift, the H$\alpha$ emission line is commonly utilized to trace the level of ongoing star formation \citep[e.g.][]{Quintero2004,Yan2006,Goto2007,French2018}. 
Since our KCWI spectral range is limited to $\lesssim$5800 \AA{}, however, this inhibits us from using H$\alpha$ as a star formation tracer.
Albeit weaker in emission and more susceptible to dust attenuation, H$\beta$ is an alternative tracer of star formation that is accessible within our KCWI spectral range and has been previously utilized to identify post-starburst systems at intermediate redshift \citep{Yan2009} and low redshift \citep{Zabludoff1996}.  
Figure~\ref{fig:EWAK} shows the distribution of H$\beta$ EW measures for our KCWI sources as a function of the best-fit ratio of A- to K-star emission (${\rm A}/{\rm K}$). 
As previously shown by various studies \citep[e.g.][]{Quintero2004, Yan2006, Yan2009}, the post-starburst population is identifiable as a spur in this space, extending to large ${\rm A}/{\rm K}$ values at very low levels of H$\beta$ emission. 
From our sample of $42$ KCWI sources with robust spectral measurements, we identify $13$ candidate post-starburst galaxies, based on a selection of 
$${\rm A}/{\rm K} > 0.3 $$
and 
$$
{\rm H}\beta~{\rm EW}~(\text{\AA}) < ({\rm A}/{\rm K}) + 0.5. 
$$
For this set of candidates, $8$ (of the $13$) are classified as UDGs.
When applied to the SDSS comparison sample, the above selection criteria yield a post-starburst fraction of $\sim 1\%$, consistent with existing measurements within the main SDSS spectroscopic sample \citep{Goto2003, Quintero2004, Yan2009}.

\begin{figure*}
\centering
\includegraphics[width=1\textwidth]{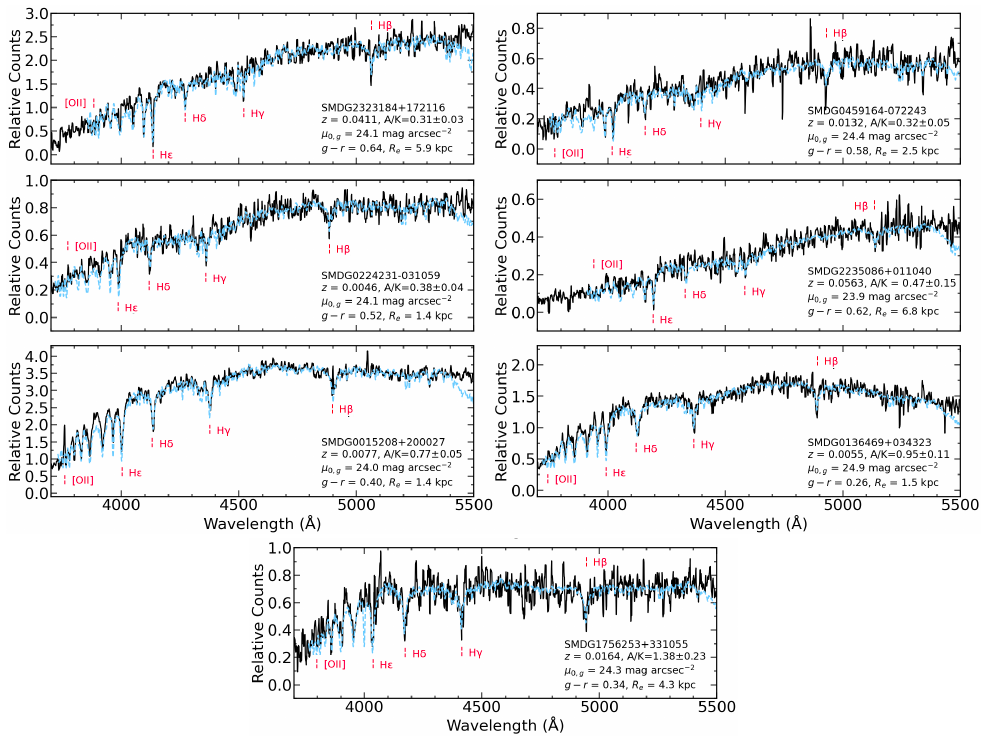}
\caption{Spectra of the $7$ post-starburst UDGs from our KCWI sample. The extracted 1-d spectra are shown in black, smoothed using an inverse-variance weighting and a kernel of 5~\AA{}, with the best-fit, composite A- and K-star template shown in blue. Each of these  sources exhibit deep Balmer absorption, consistent with a high ${\rm A}/{\rm K}$ ratio, characteristic of post-starburst galaxies. With the exception of SMDG0015208+200027, which has a modest amount of [O{\scriptsize II}] (and some H$\beta$) emission, we find very little evidence of active star formation within this population.}
\label{fig:postsbsources}
\end{figure*}

While [O{\scriptsize II}] emission can be a tracer of star formation or activity from an active galactic nucleus \citep[e.g.][]{Yan2006, Kewley2006, Yesuf2014}, we find little evidence for centrally-concentrated [O{\scriptsize II}] emission within our KCWI data cubes. Assuming star formation to be the likely source of [O{\scriptsize II}] emission, we opt to restrict our post-starburst sample to those systems with an [O{\scriptsize II}] EW of $< 6$~\AA, following the selection limits applied in previous studies of post-starburst galaxies at low and intermediate redshift \citep[e.g.][]{Balogh1999, Poggianti2009, Goto2003, Zabludoff1996}. 
Applying this [O{\scriptsize II}] selection, we remove $1$ post-starburst candidate.
This yields a final sample of $7$ post-starburst UDGs.
Figure~\ref{fig:postsbsources} shows the KCWI 1-d spectra (black) of these $7$ sources, along with the composite A- and K-star template fits (dashed blue). 
With the exception of SMDG0015208+200027, which has a modest amount of [O{\scriptsize II}] (and some H$\beta$) emission, the KCWI spectra show pronounced Balmer absorption features with no detectable sign of active star formation. 

\section{Discussion}
\label{sec:discussion}

\subsection{A Population in Transition?}
\label{subsec:population}
From our target population of $42$ sources with a measured spectroscopic redshift, we identified $13$ candidate post-starburst galaxies ($\sim 30\%$ of the total sample). 
Amongst those galaxies meeting our UDG selection criteria ($35$/$42$), we find that the post-starburst population represents a sizable fraction of $\sim20\%$ ($7$/$35$) following our ${\rm A}/{\rm K}$ ratio, H$\beta$ EW criteria, and when excluding the single source with significant [O{\scriptsize II}] emission.
As shown in \S\ref{sec:data}, our target population was largely unbiased relative to the parent SMUDGes candidate UDG sample, specifically with respect to galaxy color, such that our KCWI targets were unlikely to be biased towards post-starburst systems. 

Among the broader galaxy population at low redshift, the post-starburst fraction is typically found to be much smaller \citep[$\lesssim 1\%$ at $\mstar \gtrsim 10^{9}~\msun$,][]{Zabludoff1996, Goto2007, Yan2009, Wong2012, Pawlik2018}. 
While selection criteria for identifying post-starburst (or alternatively "transition" or "recently-quenched") galaxies can vary significantly, most studies find that post-starburst fractions increase towards lower stellar masses \citep[][but see also \citealt{Cleland2021}]{Poggianti2004, Mendel2013, Fritz2014, Aguerri2018}, with post-starbursts still only representing a few percentage of the galaxy population at $\mstar \sim 10^{8-9}~\msun$. 

This significant population of post-starburst systems in our sample suggests that we are either at a special era of UDG (and more broadly dwarf galaxy) quenching or potentially the number of post-starburst UDGs is augmented due to recycling or rejuvenation within the local UDG population, with systems moving from star-forming to quenched and back again. 
At high galaxy masses, there is little evidence for significant amounts of rejuvenation or recycling at late cosmic times, with the ages of passive systems in the local Universe generally exceeding $3$ -- $7$~Gyr \citep[e.g.][]{Worthey1994, Gallazzi2005, Daddi2005, Kriek2024}. 
For studies that do find evidence for rejuvenation, the fraction of galaxies having been rejuvenated is typically $\lesssim 10\%$ \citep[][see also \citealt{Trayford2016} for results based on simulations]{Fang2012, Cleland2021, Zhang2023}, with high rates of rejuvenation leading to over-estimates of the passive galaxy population at higher redshift relative to current observational constraints \citep{Mancini2019}.

\subsection{Environments of Post-Starburst UDGs}
\label{subsec:environ}

Independent of the potential recycling or rejuvenation within the UDG population, the post-starburst UDGs are systems where star formation has recently been quenched and as such are excellent tracers of the environment in which quenching occurs. 
Among the $7$ post-starburst UDGs in our KCWI sample, $4$ are likely satellites of more massive systems. 
For example, SMDG2323184+172116 ($z = 0.0411$) lies approximately $660$~kpc in projection from the center of a galaxy group at $z \sim 0.0428$, which is part of the Abell 2589 cluster \citep{Tully2015}. The group has an estimated mass of $\sim10^{14}~\msun$ and includes six known members listed in the 2MASS catalog \citep{Huchra2012}. 
Another case is SMDG2235086+011040 ($z = 0.056$), which resides roughly $1.2$~Mpc in projection from the center of A2457, a cluster at $z = 0.0587$ with $M_{200} \sim 5 \times 10^{14}~\msun$ and $R_{200} \sim 1.6~{\rm Mpc}$ cataloged as part of the WINGS program \citep{Moretti2017, Fasano2006, Cava2009}.
Similarly, SMDG0459164-072243 ($z = 0.0132$) is plausibly associated with a potential group containing NGC~1726 at $z \sim 0.0135$ \citep{Paturel2003, Crook2007}, residing at a projected distance of only $400$~kpc away from the massive ($\mstar \sim 10^{11.2}~\msun$) early-type galaxy \citep{Leroy2019}. 
Finally, SMDG1756253+331055 is undoubtedly a member of a group at $z \sim 0.016$ \citep{Tully2015, Tempel2018}, containing the massive disk galaxy NGC~6504 \citep{Leroy2019} that is situated less than 100~kpc away in projection.

The other $3$ post-starburst UDGs (SMDG0224231-031059, SMDG0136469+034323, and SMDG0015208+200027) are likely isolated systems. 
At $z=0.0046$, SMDG0224231-031059 is located $0.7$~Mpc in projection from NGC~936 ($z=0.0041$), which is the most massive galaxy ($\mstar \sim 10^{10.6}~\msun$) in a group containing $12$ known members \citep{Blanton2011, Makarov2011}. With an estimated group mass of $< 10^{13}~\msun$ and virial radius of $\lesssim400~{\rm kpc}$ \citep{Makarov2011}, SMDG0224231-031059 is likely unassociated with the group. 
Meanwhile, SMDG0136469+034323 ($z = 0.0055$) resides more than $1.5$~Mpc in projection from any massive ($\mstar >10^{10}~\msun$) neighbor, with a dwarf galaxy ($\mstar \sim 10^{8.5}~\msun$ at $z = 0.0055$) as its nearest neighbor ($D_{\rm proj} \sim 40~{\rm kpc}$) -- though an unlikely host.
Finally, SMDG0015208+200027 ($z = 0.0077$) also resides in relative isolation, $\gtrsim 1.5$~Mpc away (in projection) from  its nearest massive neighbor, a group of galaxies containing NGC~7817 ($\mstar \sim 10^{10.3}~\msun$ at $z = 0.0077$, \citealt{Tully2008, Tully2015, Blanton2011}). 

With the exception of the $3$ isolated (or field) post-starburst UDGs discussed above, the remainder of the quenched (H$\beta~{\rm EW} < 1~\AA$) UDG population within our KCWI sample ($\sim12$ galaxies in total) is comprised entirely of systems that are likely satellites of a more massive host or fall within the infall/backsplash region of nearby groups or clusters. 
Overall, the quenched UDG population within our KCWI sample is broadly consistent with a picture where environmental effects are predominantly responsible for suppressing star formation in the UDG population, such that the field UDG population is dominated by star-forming systems with quenched UDGs strongly biased towards massive host halos.  
This result is consistent with current observations of local UDGs and recent analyses of modern galaxy formation simulations \citep[e.g.][]{Kadowaki2021, Grishin2021, Prole2021, Jiang2019, Benavides2023}.

The 3 isolated post-starburst UDGs in our sample are potentially rare exceptions to the rule, showing evidence for quenching of UDGs outside of massive host halos (and the influence of environmental processes). 
As for some other low-density passive UDGs identified to date \citep{MD2016, Papastergis2017}, SMDG0224231-031059 is possibly a backsplash galaxy, such that it was recently quenched by the environment of NGC~936 \citep{Liao2019, Benavides2021}. 
Meanwhile, SMDG0136469+034323 and SMDG0015208+200027 reside more than $2-3~R_{\rm vir}$ away from any potential host, a distance at which  both are unlikely to be a backsplash galaxy \citep[e.g.][]{Balogh2000, Wetzel2014, Fillingham2018}. 
Along with a small number of existing objects \citep[e.g.][]{Papastergis2017, Roman2019}, these two systems present evidence that the formation of UDGs, including their quenching, may occur across a broad range of mechanisms and environments. 
For example, the existence of isolated, post-starburst UDGs may be direct evidence of the role of star formation feedback in driving size growth within the UDG population \citep[e.g.][]{DiCintio2017, Chan2018}, where a recent burst or bursts of star formation increases the size of the stellar system, while producing a significant A-star population, and then rapidly quenching the galaxy (at least temporarily). 
Given that we find all quenched UDGs (excluding the $3$ post-starburst systems, one of which does have some residual star formation) to be satellites of more massive hosts suggests that the period of star formation cessation for isolated post-starburst UDGs is short with star formation resuming quickly ($\lesssim 1$~Gyr).

\section{Summary}
\label{sec:summary}
In this work, we present a spectroscopic follow-up study of candidate ultra-diffuse galaxies selected from the SMUDGes program.
Using Keck/KCWI, we obtained spectra and measured high-confidence redshifts for 42 of 44 targets.
We investigated the properties and environments of these targets, with our main findings summarized as follows: 

\begin{itemize}[leftmargin=0.5cm]
        \item The vast majority ($35$/$42$) of candidate UDGs selected from SMUDGes were confirmed as ultra-diffuse galaxies, with the spectroscopic redshift measured from the KCWI spectra yielding sizes of $R_{e} > 1~{\rm kpc}$. \\
                
        \item Based on spectral fits and decomposition, we find post-starburst (or K+A) signatures in a large fraction of the KCWI spectroscopic sample. Approximately $\sim 30\%$ ($13$/$42$) of the sample shows minimal or no H$\beta$ emission and a prevalence of young A-type stars relative to the older K-type stellar populations. Within the confirmed UDG population, we find $20\%$ of galaxies are classified as post-starbursts, a much higher prevalence relative to that observed within the global galaxy population. \\
        
        \item Examining the environments of the post-starburst UDGs, we find that $4$ out of $7$ are satellites of more massive host galaxies, such that environmental processes (e.g.~tidal heating, ram-pressure stripping, etc.) are likely responsible for the quenching (and size growth) of these systems. The remaining $3$ post-starburst UDGs, however, are classified as isolated (or field) galaxies, such that a quenching mechanism independent of environment is likely at play in suppressing star formation (and increasing galaxy size). \\
        
\end{itemize}

\section*{Acknowledgments}

We thank the anonymous referee for valuable comments and suggestions that improved this work.
MCC acknowledges support from the National Science Foundation through grant AST-1815475. 
DCB acknowledges support via an NSF Astronomy and Astrophysics Postdoctoral Fellowship under award AST-2303800, along with support from the UC Chancellor's Postdoctoral Fellowship Program.
Much of the data presented herein were obtained at Keck Observatory, which is a private 501(c)3 non-profit organization operated as a scientific partnership among the California Institute of Technology, the University of California, and the National Aeronautics and Space Administration. The Observatory was made possible by the generous financial support of the W. M. Keck Foundation.
The authors wish to recognize and acknowledge the very significant cultural role and reverence that the summit of Maunakea has always had within the Native Hawaiian community. We are most fortunate to have the opportunity to conduct observations from this mountain.

Funding for the Sloan Digital Sky Survey V has been provided by the Alfred P. Sloan Foundation, the Heising-Simons Foundation, the National Science Foundation, and the Participating Institutions. SDSS acknowledges support and resources from the Center for High-Performance Computing at the University of Utah. SDSS telescopes are located at Apache Point Observatory, funded by the Astrophysical Research Consortium and operated by New Mexico State University, and at Las Campanas Observatory, operated by the Carnegie Institution for Science. The SDSS web site is \url{www.sdss.org}.

SDSS is managed by the Astrophysical Research Consortium for the Participating Institutions of the SDSS Collaboration, including Caltech, The Carnegie Institution for Science, Chilean National Time Allocation Committee (CNTAC) ratified researchers, The Flatiron Institute, the Gotham Participation Group, Harvard University, Heidelberg University, The Johns Hopkins University, L’Ecole polytechnique f\'{e}d\'{e}rale de Lausanne (EPFL), Leibniz-Institut f\"{u}r Astrophysik Potsdam (AIP), Max-Planck-Institut f\"{u}r Astronomie (MPIA Heidelberg), Max-Planck-Institut f\"{u}r Extraterrestrische Physik (MPE), Nanjing University, National Astronomical Observatories of China (NAOC), New Mexico State University, The Ohio State University, Pennsylvania State University, Smithsonian Astrophysical Observatory, Space Telescope Science Institute (STScI), the Stellar Astrophysics Participation Group, Universidad Nacional Aut\'{o}noma de M\'{e}xico, University of Arizona, University of Colorado Boulder, University of Illinois at Urbana-Champaign, University of Toronto, University of Utah, University of Virginia, Yale University, and Yunnan University.


\bibliographystyle{mnras}
\bibliography{Bibliography} 

\begin{thebibliography}{}
\makeatletter
\relax
\def\mn@urlcharsother{\let\do\@makeother \do\$\do\&\do\#\do\^\do\_\do\%\do\~}
\def\mn@doi{\begingroup\mn@urlcharsother \@ifnextchar [ {\mn@doi@} {\mn@doi@[]}}
\def\mn@doi@[#1]#2{\def\@tempa{#1}\ifx\@tempa\@empty \href {http://dx.doi.org/#2} {doi:#2}\else \href {http://dx.doi.org/#2} {#1}\fi \endgroup}
\def\mn@eprint#1#2{\mn@eprint@#1:#2::\@nil}
\def\mn@eprint@arXiv#1{\href {http://arxiv.org/abs/#1} {{\tt arXiv:#1}}}
\def\mn@eprint@dblp#1{\href {http://dblp.uni-trier.de/rec/bibtex/#1.xml} {dblp:#1}}
\def\mn@eprint@#1:#2:#3:#4\@nil{\def\@tempa {#1}\def\@tempb {#2}\def\@tempc {#3}\ifx \@tempc \@empty \let \@tempc \@tempb \let \@tempb \@tempa \fi \ifx \@tempb \@empty \def\@tempb {arXiv}\fi \@ifundefined {mn@eprint@\@tempb}{\@tempb:\@tempc}{\expandafter \expandafter \csname mn@eprint@\@tempb\endcsname \expandafter{\@tempc}}}

\bibitem[\protect\citeauthoryear{{Aguerri}, {Agulli}  \& {M{\'e}ndez-Abreu}}{{Aguerri} et~al.}{2018}]{Aguerri2018}
{Aguerri} J.~A.~L.,  {Agulli} I.,   {M{\'e}ndez-Abreu} J.,  2018, \mn@doi [\mnras] {10.1093/mnras/sty692}, \href {https://ui.adsabs.harvard.edu/abs/2018MNRAS.477.1921A} {477, 1921}

\bibitem[\protect\citeauthoryear{{Almeida} et~al.,}{{Almeida} et~al.}{2023}]{Almeida2023}
{Almeida} A.,  et~al., 2023, \mn@doi [\apjs] {10.3847/1538-4365/acda98}, \href {https://ui.adsabs.harvard.edu/abs/2023ApJS..267...44A} {267, 44}

\bibitem[\protect\citeauthoryear{{Amorisco} \& {Loeb}}{{Amorisco} \& {Loeb}}{2016}]{Amorisco2016}
{Amorisco} N.~C.,  {Loeb} A.,  2016, \mn@doi [\mnras] {10.1093/mnrasl/slw055}, \href {https://ui.adsabs.harvard.edu/abs/2016MNRAS.459L..51A} {459, L51}

\bibitem[\protect\citeauthoryear{{Balogh}, {Morris}, {Yee}, {Carlberg}  \& {Ellingson}}{{Balogh} et~al.}{1999}]{Balogh1999}
{Balogh} M.~L.,  {Morris} S.~L.,  {Yee} H.~K.~C.,  {Carlberg} R.~G.,   {Ellingson} E.,  1999, \mn@doi [\apj] {10.1086/308056}, \href {https://ui.adsabs.harvard.edu/abs/1999ApJ...527...54B} {527, 54}

\bibitem[\protect\citeauthoryear{{Balogh}, {Navarro}  \& {Morris}}{{Balogh} et~al.}{2000}]{Balogh2000}
{Balogh} M.~L.,  {Navarro} J.~F.,   {Morris} S.~L.,  2000, \mn@doi [\apj] {10.1086/309323}, \href {https://ui.adsabs.harvard.edu/abs/2000ApJ...540..113B} {540, 113}

\bibitem[\protect\citeauthoryear{{Baron} \& {Poznanski}}{{Baron} \& {Poznanski}}{2017}]{Baron2017}
{Baron} D.,  {Poznanski} D.,  2017, \mn@doi [\mnras] {10.1093/mnras/stw3021}, \href {https://ui.adsabs.harvard.edu/abs/2017MNRAS.465.4530B} {465, 4530}

\bibitem[\protect\citeauthoryear{{Bautista}, {Koda}, {Yagi}, {Komiyama}  \& {Yamanoi}}{{Bautista} et~al.}{2023}]{Bautista2023}
{Bautista} J. M.~G.,  {Koda} J.,  {Yagi} M.,  {Komiyama} Y.,   {Yamanoi} H.,  2023, \mn@doi [\apjs] {10.3847/1538-4365/acd3e7}, \href {https://ui.adsabs.harvard.edu/abs/2023ApJS..267...10B} {267, 10}

\bibitem[\protect\citeauthoryear{{Benavides} et~al.,}{{Benavides} et~al.}{2021}]{Benavides2021}
{Benavides} J.~A.,  et~al., 2021, \mn@doi [Nature Astronomy] {10.1038/s41550-021-01458-1}, \href {https://ui.adsabs.harvard.edu/abs/2021NatAs...5.1255B} {5, 1255}

\bibitem[\protect\citeauthoryear{{Benavides}, {Sales}, {Abadi}, {Marinacci}, {Vogelsberger}  \& {Hernquist}}{{Benavides} et~al.}{2023}]{Benavides2023}
{Benavides} J.~A.,  {Sales} L.~V.,  {Abadi} M.~G.,  {Marinacci} F.,  {Vogelsberger} M.,   {Hernquist} L.,  2023, \mn@doi [\mnras] {10.1093/mnras/stad1053}, \href {https://ui.adsabs.harvard.edu/abs/2023MNRAS.522.1033B} {522, 1033}

\bibitem[\protect\citeauthoryear{{Bennet}, {Sand}, {Zaritsky}, {Crnojevi{\'c}}, {Spekkens}  \& {Karunakaran}}{{Bennet} et~al.}{2018}]{Bennet2018}
{Bennet} P.,  {Sand} D.~J.,  {Zaritsky} D.,  {Crnojevi{\'c}} D.,  {Spekkens} K.,   {Karunakaran} A.,  2018, \mn@doi [\apjl] {10.3847/2041-8213/aadedf}, \href {https://ui.adsabs.harvard.edu/abs/2018ApJ...866L..11B} {866, L11}

\bibitem[\protect\citeauthoryear{{Bezanson} et~al.,}{{Bezanson} et~al.}{2022}]{Bezanson2022}
{Bezanson} R.,  et~al., 2022, \mn@doi [\apj] {10.3847/1538-4357/ac3dfa}, \href {https://ui.adsabs.harvard.edu/abs/2022ApJ...925..153B} {925, 153}

\bibitem[\protect\citeauthoryear{{Blanton}, {Kazin}, {Muna}, {Weaver}  \& {Price-Whelan}}{{Blanton} et~al.}{2011}]{Blanton2011}
{Blanton} M.~R.,  {Kazin} E.,  {Muna} D.,  {Weaver} B.~A.,   {Price-Whelan} A.,  2011, \mn@doi [\aj] {10.1088/0004-6256/142/1/31}, \href {https://ui.adsabs.harvard.edu/abs/2011AJ....142...31B} {142, 31}

\bibitem[\protect\citeauthoryear{{Bothun}, {Impey}  \& {Malin}}{{Bothun} et~al.}{1991}]{Bothun1991}
{Bothun} G.~D.,  {Impey} C.~D.,   {Malin} D.~F.,  1991, \mn@doi [\apj] {10.1086/170290}, \href {https://ui.adsabs.harvard.edu/abs/1991ApJ...376..404B} {376, 404}

\bibitem[\protect\citeauthoryear{{Brinchmann}, {Charlot}, {White}, {Tremonti}, {Kauffmann}, {Heckman}  \& {Brinkmann}}{{Brinchmann} et~al.}{2004}]{Brinchmann2004}
{Brinchmann} J.,  {Charlot} S.,  {White} S.~D.~M.,  {Tremonti} C.,  {Kauffmann} G.,  {Heckman} T.,   {Brinkmann} J.,  2004, \mn@doi [\mnras] {10.1111/j.1365-2966.2004.07881.x}, \href {https://ui.adsabs.harvard.edu/abs/2004MNRAS.351.1151B} {351, 1151}

\bibitem[\protect\citeauthoryear{{Carleton}, {Errani}, {Cooper}, {Kaplinghat}, {Pe{\~n}arrubia}  \& {Guo}}{{Carleton} et~al.}{2019}]{Carleton2019}
{Carleton} T.,  {Errani} R.,  {Cooper} M.,  {Kaplinghat} M.,  {Pe{\~n}arrubia} J.,   {Guo} Y.,  2019, \mn@doi [\mnras] {10.1093/mnras/stz383}, \href {https://ui.adsabs.harvard.edu/abs/2019MNRAS.485..382C} {485, 382}

\bibitem[\protect\citeauthoryear{{Carleton} et~al.,}{{Carleton} et~al.}{2023}]{Carleton2023}
{Carleton} T.,  et~al., 2023, \mn@doi [\apj] {10.3847/1538-4357/ace343}, \href {https://ui.adsabs.harvard.edu/abs/2023ApJ...953...83C} {953, 83}

\bibitem[\protect\citeauthoryear{{Carnall}, {McLure}, {Dunlop}  \& {Dav{\'e}}}{{Carnall} et~al.}{2018}]{Carnall2018}
{Carnall} A.~C.,  {McLure} R.~J.,  {Dunlop} J.~S.,   {Dav{\'e}} R.,  2018, \mn@doi [\mnras] {10.1093/mnras/sty2169}, \href {https://ui.adsabs.harvard.edu/abs/2018MNRAS.480.4379C} {480, 4379}

\bibitem[\protect\citeauthoryear{{Cava} et~al.,}{{Cava} et~al.}{2009}]{Cava2009}
{Cava} A.,  et~al., 2009, \mn@doi [\aap] {10.1051/0004-6361:200810997}, \href {https://ui.adsabs.harvard.edu/abs/2009A&A...495..707C} {495, 707}

\bibitem[\protect\citeauthoryear{{Chan}, {Kere{\v{s}}}, {Wetzel}, {Hopkins}, {Faucher-Gigu{\`e}re}, {El-Badry}, {Garrison-Kimmel}  \& {Boylan-Kolchin}}{{Chan} et~al.}{2018}]{Chan2018}
{Chan} T.~K.,  {Kere{\v{s}}} D.,  {Wetzel} A.,  {Hopkins} P.~F.,  {Faucher-Gigu{\`e}re} C.~A.,  {El-Badry} K.,  {Garrison-Kimmel} S.,   {Boylan-Kolchin} M.,  2018, \mn@doi [\mnras] {10.1093/mnras/sty1153}, \href {https://ui.adsabs.harvard.edu/abs/2018MNRAS.478..906C} {478, 906}

\bibitem[\protect\citeauthoryear{{Cleland} \& {McGee}}{{Cleland} \& {McGee}}{2021}]{Cleland2021}
{Cleland} C.,  {McGee} S.~L.,  2021, \mn@doi [\mnras] {10.1093/mnras/staa3267}, \href {https://ui.adsabs.harvard.edu/abs/2021MNRAS.500..590C} {500, 590}

\bibitem[\protect\citeauthoryear{{Crook}, {Huchra}, {Martimbeau}, {Masters}, {Jarrett}  \& {Macri}}{{Crook} et~al.}{2007}]{Crook2007}
{Crook} A.~C.,  {Huchra} J.~P.,  {Martimbeau} N.,  {Masters} K.~L.,  {Jarrett} T.,   {Macri} L.~M.,  2007, \mn@doi [\apj] {10.1086/510201}, \href {https://ui.adsabs.harvard.edu/abs/2007ApJ...655..790C} {655, 790}

\bibitem[\protect\citeauthoryear{{Daddi} et~al.,}{{Daddi} et~al.}{2005}]{Daddi2005}
{Daddi} E.,  et~al., 2005, \mn@doi [\apj] {10.1086/430104}, \href {https://ui.adsabs.harvard.edu/abs/2005ApJ...626..680D} {626, 680}

\bibitem[\protect\citeauthoryear{{Dey} et~al.,}{{Dey} et~al.}{2016}]{Dey2016}
{Dey} A.,  et~al., 2016, in {Evans} C.~J.,  {Simard} L.,   {Takami} H.,  eds,  Society of Photo-Optical Instrumentation Engineers (SPIE) Conference Series Vol. 9908, Ground-based and Airborne Instrumentation for Astronomy VI. p. 99082C, \mn@doi{10.1117/12.2231488}

\bibitem[\protect\citeauthoryear{{Dey} et~al.,}{{Dey} et~al.}{2019}]{dey2019}
{Dey} A.,  et~al., 2019, \mn@doi [\aj] {10.3847/1538-3881/ab089d}, \href {https://ui.adsabs.harvard.edu/abs/2019AJ....157..168D} {157, 168}

\bibitem[\protect\citeauthoryear{{Di Cintio}, {Brook}, {Dutton}, {Macci{\`o}}, {Obreja}  \& {Dekel}}{{Di Cintio} et~al.}{2017}]{DiCintio2017}
{Di Cintio} A.,  {Brook} C.~B.,  {Dutton} A.~A.,  {Macci{\`o}} A.~V.,  {Obreja} A.,   {Dekel} A.,  2017, \mn@doi [\mnras] {10.1093/mnrasl/slw210}, \href {https://ui.adsabs.harvard.edu/abs/2017MNRAS.466L...1D} {466, L1}

\bibitem[\protect\citeauthoryear{{Dressler} \& {Gunn}}{{Dressler} \& {Gunn}}{1983}]{Dressler1983}
{Dressler} A.,  {Gunn} J.~E.,  1983, \mn@doi [\apj] {10.1086/161093}, \href {https://ui.adsabs.harvard.edu/abs/1983ApJ...270....7D} {270, 7}

\bibitem[\protect\citeauthoryear{{Eigenthaler} et~al.,}{{Eigenthaler} et~al.}{2018}]{Eigenthaler2018}
{Eigenthaler} P.,  et~al., 2018, \mn@doi [\apj] {10.3847/1538-4357/aaab60}, \href {https://ui.adsabs.harvard.edu/abs/2018ApJ...855..142E} {855, 142}

\bibitem[\protect\citeauthoryear{{Eisenstein} et~al.,}{{Eisenstein} et~al.}{2003}]{Eisenstein2003}
{Eisenstein} D.~J.,  et~al., 2003, \mn@doi [\apj] {10.1086/346233}, \href {https://ui.adsabs.harvard.edu/abs/2003ApJ...585..694E} {585, 694}

\bibitem[\protect\citeauthoryear{{Fang}, {Faber}, {Salim}, {Graves}  \& {Rich}}{{Fang} et~al.}{2012}]{Fang2012}
{Fang} J.~J.,  {Faber} S.~M.,  {Salim} S.,  {Graves} G.~J.,   {Rich} R.~M.,  2012, \mn@doi [\apj] {10.1088/0004-637X/761/1/23}, \href {https://ui.adsabs.harvard.edu/abs/2012ApJ...761...23F} {761, 23}

\bibitem[\protect\citeauthoryear{{Fasano} et~al.,}{{Fasano} et~al.}{2006}]{Fasano2006}
{Fasano} G.,  et~al., 2006, \mn@doi [\aap] {10.1051/0004-6361:20053816}, \href {https://ui.adsabs.harvard.edu/abs/2006A&A...445..805F} {445, 805}

\bibitem[\protect\citeauthoryear{{Fielder}, {Jones}, {Sand}, {Bennet}, {Crnojevic}, {Karunakaran}, {Mutlu-Pakdil}  \& {Spekkens}}{{Fielder} et~al.}{2024}]{Fielder2024}
{Fielder} C.,  {Jones} M.,  {Sand} D.,  {Bennet} P.,  {Crnojevic} D.,  {Karunakaran} A.,  {Mutlu-Pakdil} B.,   {Spekkens} K.,  2024, \mn@doi [arXiv e-prints] {10.48550/arXiv.2401.01931}, \href {https://ui.adsabs.harvard.edu/abs/2024arXiv240101931F} {p. arXiv:2401.01931}

\bibitem[\protect\citeauthoryear{{Fillingham}, {Cooper}, {Boylan-Kolchin}, {Bullock}, {Garrison-Kimmel}  \& {Wheeler}}{{Fillingham} et~al.}{2018}]{Fillingham2018}
{Fillingham} S.~P.,  {Cooper} M.~C.,  {Boylan-Kolchin} M.,  {Bullock} J.~S.,  {Garrison-Kimmel} S.,   {Wheeler} C.,  2018, \mn@doi [\mnras] {10.1093/mnras/sty958}, \href {https://ui.adsabs.harvard.edu/abs/2018MNRAS.477.4491F} {477, 4491}

\bibitem[\protect\citeauthoryear{{Flaugher} et~al.,}{{Flaugher} et~al.}{2015}]{Flaugher2015}
{Flaugher} B.,  et~al., 2015, \mn@doi [\aj] {10.1088/0004-6256/150/5/150}, \href {https://ui.adsabs.harvard.edu/abs/2015AJ....150..150F} {150, 150}

\bibitem[\protect\citeauthoryear{{Forbes} et~al.,}{{Forbes} et~al.}{2023}]{Forbes2023}
{Forbes} D.~A.,  et~al., 2023, \mn@doi [\mnras] {10.1093/mnrasl/slad101}, \href {https://ui.adsabs.harvard.edu/abs/2023MNRAS.525L..93F} {525, L93}

\bibitem[\protect\citeauthoryear{{French}, {Yang}, {Zabludoff}  \& {Tremonti}}{{French} et~al.}{2018}]{French2018}
{French} K.~D.,  {Yang} Y.,  {Zabludoff} A.~I.,   {Tremonti} C.~A.,  2018, \mn@doi [\apj] {10.3847/1538-4357/aacb2d}, \href {https://ui.adsabs.harvard.edu/abs/2018ApJ...862....2F} {862, 2}

\bibitem[\protect\citeauthoryear{{Freundlich}, {Dekel}, {Jiang}, {Ishai}, {Cornuault}, {Lapiner}, {Dutton}  \& {Macci{\`o}}}{{Freundlich} et~al.}{2020}]{Freundlich2020}
{Freundlich} J.,  {Dekel} A.,  {Jiang} F.,  {Ishai} G.,  {Cornuault} N.,  {Lapiner} S.,  {Dutton} A.~A.,   {Macci{\`o}} A.~V.,  2020, \mn@doi [\mnras] {10.1093/mnras/stz3306}, \href {https://ui.adsabs.harvard.edu/abs/2020MNRAS.491.4523F} {491, 4523}

\bibitem[\protect\citeauthoryear{{Fritz} et~al.,}{{Fritz} et~al.}{2014}]{Fritz2014}
{Fritz} J.,  et~al., 2014, \mn@doi [\aap] {10.1051/0004-6361/201323138}, \href {https://ui.adsabs.harvard.edu/abs/2014A&A...566A..32F} {566, A32}

\bibitem[\protect\citeauthoryear{{Gallazzi}, {Charlot}, {Brinchmann}, {White}  \& {Tremonti}}{{Gallazzi} et~al.}{2005}]{Gallazzi2005}
{Gallazzi} A.,  {Charlot} S.,  {Brinchmann} J.,  {White} S. D.~M.,   {Tremonti} C.~A.,  2005, \mn@doi [\mnras] {10.1111/j.1365-2966.2005.09321.x}, \href {https://ui.adsabs.harvard.edu/abs/2005MNRAS.362...41G} {362, 41}

\bibitem[\protect\citeauthoryear{{Gannon} et~al.,}{{Gannon} et~al.}{2021}]{Gannon2021}
{Gannon} J.~S.,  et~al., 2021, \mn@doi [\mnras] {10.1093/mnras/stab277}, \href {https://ui.adsabs.harvard.edu/abs/2021MNRAS.502.3144G} {502, 3144}

\bibitem[\protect\citeauthoryear{{Gannon} et~al.,}{{Gannon} et~al.}{2022}]{Gannon2022}
{Gannon} J.~S.,  et~al., 2022, \mn@doi [\mnras] {10.1093/mnras/stab3297}, \href {https://ui.adsabs.harvard.edu/abs/2022MNRAS.510..946G} {510, 946}

\bibitem[\protect\citeauthoryear{{Geha}, {Blanton}, {Yan}  \& {Tinker}}{{Geha} et~al.}{2012}]{Geha2012}
{Geha} M.,  {Blanton} M.~R.,  {Yan} R.,   {Tinker} J.~L.,  2012, \mn@doi [\apj] {10.1088/0004-637X/757/1/85}, \href {https://ui.adsabs.harvard.edu/abs/2012ApJ...757...85G} {757, 85}

\bibitem[\protect\citeauthoryear{{Gil de Paz} et~al.,}{{Gil de Paz} et~al.}{2007}]{GildePaz2007}
{Gil de Paz} A.,  et~al., 2007, \mn@doi [\apjs] {10.1086/516636}, \href {https://ui.adsabs.harvard.edu/abs/2007ApJS..173..185G} {173, 185}

\bibitem[\protect\citeauthoryear{{Goto}}{{Goto}}{2007}]{Goto2007}
{Goto} T.,  2007, \mn@doi [\mnras] {10.1111/j.1365-2966.2007.12227.x}, \href {https://ui.adsabs.harvard.edu/abs/2007MNRAS.381..187G} {381, 187}

\bibitem[\protect\citeauthoryear{{Goto} et~al.,}{{Goto} et~al.}{2003}]{Goto2003}
{Goto} T.,  et~al., 2003, \mn@doi [\pasj] {10.1093/pasj/55.4.771}, \href {https://ui.adsabs.harvard.edu/abs/2003PASJ...55..771G} {55, 771}

\bibitem[\protect\citeauthoryear{{Green}}{{Green}}{2018}]{Green2018}
{Green} G.~M.,  2018, \mn@doi [The Journal of Open Source Software] {10.21105/joss.00695}, \href {https://ui.adsabs.harvard.edu/abs/2018JOSS....3..695G} {3, 695}

\bibitem[\protect\citeauthoryear{{Grishin}, {Chilingarian}, {Afanasiev}, {Fabricant}, {Katkov}, {Moran}  \& {Yagi}}{{Grishin} et~al.}{2021}]{Grishin2021}
{Grishin} K.~A.,  {Chilingarian} I.~V.,  {Afanasiev} A.~V.,  {Fabricant} D.,  {Katkov} I.~Y.,  {Moran} S.,   {Yagi} M.,  2021, \mn@doi [Nature Astronomy] {10.1038/s41550-021-01470-5}, \href {https://ui.adsabs.harvard.edu/abs/2021NatAs...5.1308G} {5, 1308}

\bibitem[\protect\citeauthoryear{{Huchra} et~al.,}{{Huchra} et~al.}{2012}]{Huchra2012}
{Huchra} J.~P.,  et~al., 2012, \mn@doi [\apjs] {10.1088/0067-0049/199/2/26}, \href {https://ui.adsabs.harvard.edu/abs/2012ApJS..199...26H} {199, 26}

\bibitem[\protect\citeauthoryear{{Hutchinson} et~al.,}{{Hutchinson} et~al.}{2016}]{Hutchinson2016}
{Hutchinson} T.~A.,  et~al., 2016, \mn@doi [\aj] {10.3847/0004-6256/152/6/205}, \href {https://ui.adsabs.harvard.edu/abs/2016AJ....152..205H} {152, 205}

\bibitem[\protect\citeauthoryear{{Impey} \& {Bothun}}{{Impey} \& {Bothun}}{1997}]{Impey1997}
{Impey} C.,  {Bothun} G.,  1997, \mn@doi [\araa] {10.1146/annurev.astro.35.1.267}, \href {https://ui.adsabs.harvard.edu/abs/1997ARA&A..35..267I} {35, 267}

\bibitem[\protect\citeauthoryear{Jiang, Dekel, Freundlich, Romanowsky, Dutton, Macci{\`o}  \& Di~Cintio}{Jiang et~al.}{2019}]{Jiang2019}
Jiang F.,  Dekel A.,  Freundlich J.,  Romanowsky A.~J.,  Dutton A.~A.,  Macci{\`o} A.~V.,   Di~Cintio A.,  2019, \mn@doi [Monthly Notices of the Royal Astronomical Society] {10.1093/mnras/stz1499}, 487, 5272

\bibitem[\protect\citeauthoryear{{Johnson}, {Leja}, {Conroy}  \& {Speagle}}{{Johnson} et~al.}{2021}]{Johnson2021}
{Johnson} B.~D.,  {Leja} J.,  {Conroy} C.,   {Speagle} J.~S.,  2021, \mn@doi [\apjs] {10.3847/1538-4365/abef67}, \href {https://ui.adsabs.harvard.edu/abs/2021ApJS..254...22J} {254, 22}

\bibitem[\protect\citeauthoryear{{Jones}, {Bennet}, {Mutlu-Pakdil}, {Sand}, {Spekkens}, {Crnojevi{\'c}}, {Karunakaran}  \& {Zaritsky}}{{Jones} et~al.}{2021}]{Jones2021}
{Jones} M.~G.,  {Bennet} P.,  {Mutlu-Pakdil} B.,  {Sand} D.~J.,  {Spekkens} K.,  {Crnojevi{\'c}} D.,  {Karunakaran} A.,   {Zaritsky} D.,  2021, \mn@doi [\apj] {10.3847/1538-4357/ac0975}, \href {https://ui.adsabs.harvard.edu/abs/2021ApJ...919...72J} {919, 72}

\bibitem[\protect\citeauthoryear{{Kadowaki}, {Zaritsky}, {Donnerstein}, {RS}, {Karunakaran}  \& {Spekkens}}{{Kadowaki} et~al.}{2021}]{Kadowaki2021}
{Kadowaki} J.,  {Zaritsky} D.,  {Donnerstein} R.~L.,  {RS} P.,  {Karunakaran} A.,   {Spekkens} K.,  2021, \mn@doi [\apj] {10.3847/1538-4357/ac2948}, \href {https://ui.adsabs.harvard.edu/abs/2021ApJ...923..257K} {923, 257}

\bibitem[\protect\citeauthoryear{{Kauffmann} et~al.,}{{Kauffmann} et~al.}{2003}]{Kauffmann2003}
{Kauffmann} G.,  et~al., 2003, \mn@doi [\mnras] {10.1046/j.1365-8711.2003.06291.x}, \href {https://ui.adsabs.harvard.edu/abs/2003MNRAS.341...33K} {341, 33}

\bibitem[\protect\citeauthoryear{{Kewley}, {Groves}, {Kauffmann}  \& {Heckman}}{{Kewley} et~al.}{2006}]{Kewley2006}
{Kewley} L.~J.,  {Groves} B.,  {Kauffmann} G.,   {Heckman} T.,  2006, \mn@doi [\mnras] {10.1111/j.1365-2966.2006.10859.x}, \href {https://ui.adsabs.harvard.edu/abs/2006MNRAS.372..961K} {372, 961}

\bibitem[\protect\citeauthoryear{{Koda}, {Yagi}, {Yamanoi}  \& {Komiyama}}{{Koda} et~al.}{2015}]{Koda2015}
{Koda} J.,  {Yagi} M.,  {Yamanoi} H.,   {Komiyama} Y.,  2015, \mn@doi [\apjl] {10.1088/2041-8205/807/1/L2}, \href {https://ui.adsabs.harvard.edu/abs/2015ApJ...807L...2K} {807, L2}

\bibitem[\protect\citeauthoryear{{Kriek} et~al.,}{{Kriek} et~al.}{2024}]{Kriek2024}
{Kriek} M.,  et~al., 2024, \mn@doi [\apj] {10.3847/1538-4357/ad2df9}, \href {https://ui.adsabs.harvard.edu/abs/2024ApJ...966...36K} {966, 36}

\bibitem[\protect\citeauthoryear{{Leisman} et~al.,}{{Leisman} et~al.}{2017}]{Leisman2017}
{Leisman} L.,  et~al., 2017, \mn@doi [\apj] {10.3847/1538-4357/aa7575}, \href {https://ui.adsabs.harvard.edu/abs/2017ApJ...842..133L} {842, 133}

\bibitem[\protect\citeauthoryear{{Leroy} et~al.,}{{Leroy} et~al.}{2019}]{Leroy2019}
{Leroy} A.~K.,  et~al., 2019, \mn@doi [\apjs] {10.3847/1538-4365/ab3925}, \href {https://ui.adsabs.harvard.edu/abs/2019ApJS..244...24L} {244, 24}

\bibitem[\protect\citeauthoryear{{Liao} et~al.,}{{Liao} et~al.}{2019}]{Liao2019}
{Liao} S.,  et~al., 2019, \mn@doi [\mnras] {10.1093/mnras/stz2969}, \href {https://ui.adsabs.harvard.edu/abs/2019MNRAS.490.5182L} {490, 5182}

\bibitem[\protect\citeauthoryear{{Makarov} \& {Karachentsev}}{{Makarov} \& {Karachentsev}}{2011}]{Makarov2011}
{Makarov} D.,  {Karachentsev} I.,  2011, \mn@doi [\mnras] {10.1111/j.1365-2966.2010.18071.x}, \href {https://ui.adsabs.harvard.edu/abs/2011MNRAS.412.2498M} {412, 2498}

\bibitem[\protect\citeauthoryear{{Mancini} et~al.,}{{Mancini} et~al.}{2019}]{Mancini2019}
{Mancini} C.,  et~al., 2019, \mn@doi [\mnras] {10.1093/mnras/stz2130}, \href {https://ui.adsabs.harvard.edu/abs/2019MNRAS.489.1265M} {489, 1265}

\bibitem[\protect\citeauthoryear{{Marleau} et~al.,}{{Marleau} et~al.}{2021}]{Marleau2021}
{Marleau} F.~R.,  et~al., 2021, \mn@doi [\aap] {10.1051/0004-6361/202141432}, \href {https://ui.adsabs.harvard.edu/abs/2021A&A...654A.105M} {654, A105}

\bibitem[\protect\citeauthoryear{{Martin}, {Moore}, {Morrissey}, {Matuszewski}, {Rahman}, {Adkins}  \& {Epps}}{{Martin} et~al.}{2010}]{Martin2010}
{Martin} C.,  {Moore} A.,  {Morrissey} P.,  {Matuszewski} M.,  {Rahman} S.,  {Adkins} S.,   {Epps} H.,  2010, in {McLean} I.~S.,  {Ramsay} S.~K.,   {Takami} H.,  eds,  Society of Photo-Optical Instrumentation Engineers (SPIE) Conference Series Vol. 7735, Ground-based and Airborne Instrumentation for Astronomy III. p. 77350M, \mn@doi{10.1117/12.858227}

\bibitem[\protect\citeauthoryear{{Mart{\'\i}nez-Delgado} et~al.,}{{Mart{\'\i}nez-Delgado} et~al.}{2016}]{MD2016}
{Mart{\'\i}nez-Delgado} D.,  et~al., 2016, \mn@doi [\aj] {10.3847/0004-6256/151/4/96}, \href {https://ui.adsabs.harvard.edu/abs/2016AJ....151...96M} {151, 96}

\bibitem[\protect\citeauthoryear{{Mendel}, {Simard}, {Ellison}  \& {Patton}}{{Mendel} et~al.}{2013}]{Mendel2013}
{Mendel} J.~T.,  {Simard} L.,  {Ellison} S.~L.,   {Patton} D.~R.,  2013, \mn@doi [\mnras] {10.1093/mnras/sts489}, \href {https://ui.adsabs.harvard.edu/abs/2013MNRAS.429.2212M} {429, 2212}

\bibitem[\protect\citeauthoryear{{Meusinger}, {Br{\"u}necke}, {Schalldach}  \& {in der Au}}{{Meusinger} et~al.}{2017}]{Meusinger2017}
{Meusinger} H.,  {Br{\"u}necke} J.,  {Schalldach} P.,   {in der Au} A.,  2017, \mn@doi [\aap] {10.1051/0004-6361/201629139}, \href {https://ui.adsabs.harvard.edu/abs/2017A&A...597A.134M} {597, A134}

\bibitem[\protect\citeauthoryear{{Mihos} et~al.,}{{Mihos} et~al.}{2015}]{Mihos2015}
{Mihos} J.~C.,  et~al., 2015, \mn@doi [\apjl] {10.1088/2041-8205/809/2/L21}, \href {https://ui.adsabs.harvard.edu/abs/2015ApJ...809L..21M} {809, L21}

\bibitem[\protect\citeauthoryear{{Moretti} et~al.,}{{Moretti} et~al.}{2017}]{Moretti2017}
{Moretti} A.,  et~al., 2017, \mn@doi [\aap] {10.1051/0004-6361/201630030}, \href {https://ui.adsabs.harvard.edu/abs/2017A&A...599A..81M} {599, A81}

\bibitem[\protect\citeauthoryear{{Morrissey} et~al.,}{{Morrissey} et~al.}{2018}]{Morrissey2018}
{Morrissey} P.,  et~al., 2018, \mn@doi [\apj] {10.3847/1538-4357/aad597}, \href {https://ui.adsabs.harvard.edu/abs/2018ApJ...864...93M} {864, 93}

\bibitem[\protect\citeauthoryear{{Oke} \& {Gunn}}{{Oke} \& {Gunn}}{1983}]{Oke1983}
{Oke} J.~B.,  {Gunn} J.~E.,  1983, \mn@doi [\apj] {10.1086/160817}, \href {https://ui.adsabs.harvard.edu/abs/1983ApJ...266..713O} {266, 713}

\bibitem[\protect\citeauthoryear{{Papastergis}, {Adams}  \& {Romanowsky}}{{Papastergis} et~al.}{2017}]{Papastergis2017}
{Papastergis} E.,  {Adams} E.~A.~K.,   {Romanowsky} A.~J.,  2017, \mn@doi [\aap] {10.1051/0004-6361/201730795}, \href {https://ui.adsabs.harvard.edu/abs/2017A&A...601L..10P} {601, L10}

\bibitem[\protect\citeauthoryear{{Paturel}, {Petit}, {Prugniel}, {Theureau}, {Rousseau}, {Brouty}, {Dubois}  \& {Cambr{\'e}sy}}{{Paturel} et~al.}{2003}]{Paturel2003}
{Paturel} G.,  {Petit} C.,  {Prugniel} P.,  {Theureau} G.,  {Rousseau} J.,  {Brouty} M.,  {Dubois} P.,   {Cambr{\'e}sy} L.,  2003, \mn@doi [\aap] {10.1051/0004-6361:20031411}, \href {https://ui.adsabs.harvard.edu/abs/2003A&A...412...45P} {412, 45}

\bibitem[\protect\citeauthoryear{{Pawlik} et~al.,}{{Pawlik} et~al.}{2018}]{Pawlik2018}
{Pawlik} M.~M.,  et~al., 2018, \mn@doi [\mnras] {10.1093/mnras/sty589}, \href {https://ui.adsabs.harvard.edu/abs/2018MNRAS.477.1708P} {477, 1708}

\bibitem[\protect\citeauthoryear{{Poggianti}, {Bridges}, {Komiyama}, {Yagi}, {Carter}, {Mobasher}, {Okamura}  \& {Kashikawa}}{{Poggianti} et~al.}{2004}]{Poggianti2004}
{Poggianti} B.~M.,  {Bridges} T.~J.,  {Komiyama} Y.,  {Yagi} M.,  {Carter} D.,  {Mobasher} B.,  {Okamura} S.,   {Kashikawa} N.,  2004, \mn@doi [\apj] {10.1086/380195}, \href {https://ui.adsabs.harvard.edu/abs/2004ApJ...601..197P} {601, 197}

\bibitem[\protect\citeauthoryear{{Poggianti} et~al.,}{{Poggianti} et~al.}{2009}]{Poggianti2009}
{Poggianti} B.~M.,  et~al., 2009, \mn@doi [\apj] {10.1088/0004-637X/693/1/112}, \href {https://ui.adsabs.harvard.edu/abs/2009ApJ...693..112P} {693, 112}

\bibitem[\protect\citeauthoryear{{Prole}, {van der Burg}, {Hilker}  \& {Spitler}}{{Prole} et~al.}{2021}]{Prole2021}
{Prole} D.~J.,  {van der Burg} R.~F.~J.,  {Hilker} M.,   {Spitler} L.~R.,  2021, \mn@doi [\mnras] {10.1093/mnras/staa3296}, \href {https://ui.adsabs.harvard.edu/abs/2021MNRAS.500.2049P} {500, 2049}

\bibitem[\protect\citeauthoryear{{Quintero} et~al.,}{{Quintero} et~al.}{2004}]{Quintero2004}
{Quintero} A.~D.,  et~al., 2004, \mn@doi [\apj] {10.1086/380601}, \href {https://ui.adsabs.harvard.edu/abs/2004ApJ...602..190Q} {602, 190}

\bibitem[\protect\citeauthoryear{{Roediger} \& {Courteau}}{{Roediger} \& {Courteau}}{2015}]{Roediger2015}
{Roediger} J.~C.,  {Courteau} S.,  2015, \mn@doi [\mnras] {10.1093/mnras/stv1499}, \href {https://ui.adsabs.harvard.edu/abs/2015MNRAS.452.3209R} {452, 3209}

\bibitem[\protect\citeauthoryear{{Rom{\'a}n} \& {Trujillo}}{{Rom{\'a}n} \& {Trujillo}}{2017a}]{Roman2017a}
{Rom{\'a}n} J.,  {Trujillo} I.,  2017a, \mn@doi [\mnras] {10.1093/mnras/stx438}, \href {https://ui.adsabs.harvard.edu/abs/2017MNRAS.468..703R} {468, 703}

\bibitem[\protect\citeauthoryear{{Rom{\'a}n} \& {Trujillo}}{{Rom{\'a}n} \& {Trujillo}}{2017b}]{Roman2017b}
{Rom{\'a}n} J.,  {Trujillo} I.,  2017b, \mn@doi [\mnras] {10.1093/mnras/stx694}, \href {https://ui.adsabs.harvard.edu/abs/2017MNRAS.468.4039R} {468, 4039}

\bibitem[\protect\citeauthoryear{{Rom{\'a}n}, {Beasley}, {Ruiz-Lara}  \& {Valls-Gabaud}}{{Rom{\'a}n} et~al.}{2019}]{Roman2019}
{Rom{\'a}n} J.,  {Beasley} M.~A.,  {Ruiz-Lara} T.,   {Valls-Gabaud} D.,  2019, \mn@doi [\mnras] {10.1093/mnras/stz835}, \href {https://ui.adsabs.harvard.edu/abs/2019MNRAS.486..823R} {486, 823}

\bibitem[\protect\citeauthoryear{{Rong}, {Zhu}, {Johnston}, {Zhang}, {Cao}, {Puzia}  \& {Galaz}}{{Rong} et~al.}{2020}]{Rong2020a}
{Rong} Y.,  {Zhu} K.,  {Johnston} E.~J.,  {Zhang} H.-X.,  {Cao} T.,  {Puzia} T.~H.,   {Galaz} G.,  2020, \mn@doi [\apjl] {10.3847/2041-8213/aba8aa}, \href {https://ui.adsabs.harvard.edu/abs/2020ApJ...899L..12R} {899, L12}

\bibitem[\protect\citeauthoryear{{Sales}, {Navarro}, {Pe{\~n}afiel}, {Peng}, {Lim}  \& {Hernquist}}{{Sales} et~al.}{2020}]{Sales2020}
{Sales} L.~V.,  {Navarro} J.~F.,  {Pe{\~n}afiel} L.,  {Peng} E.~W.,  {Lim} S.,   {Hernquist} L.,  2020, \mn@doi [\mnras] {10.1093/mnras/staa854}, \href {https://ui.adsabs.harvard.edu/abs/2020MNRAS.494.1848S} {494, 1848}

\bibitem[\protect\citeauthoryear{{S{\'a}nchez-Bl{\'a}zquez} et~al.,}{{S{\'a}nchez-Bl{\'a}zquez} et~al.}{2006}]{Sanchez2006}
{S{\'a}nchez-Bl{\'a}zquez} P.,  et~al., 2006, \mn@doi [\mnras] {10.1111/j.1365-2966.2006.10699.x}, \href {https://ui.adsabs.harvard.edu/abs/2006MNRAS.371..703S} {371, 703}

\bibitem[\protect\citeauthoryear{{Sandage} \& {Binggeli}}{{Sandage} \& {Binggeli}}{1984}]{Sandage1984}
{Sandage} A.,  {Binggeli} B.,  1984, \mn@doi [\aj] {10.1086/113588}, \href {https://ui.adsabs.harvard.edu/abs/1984AJ.....89..919S} {89, 919}

\bibitem[\protect\citeauthoryear{{Schlegel}, {Finkbeiner}  \& {Davis}}{{Schlegel} et~al.}{1998}]{Schlegel1998}
{Schlegel} D.~J.,  {Finkbeiner} D.~P.,   {Davis} M.,  1998, \mn@doi [\apj] {10.1086/305772}, \href {https://ui.adsabs.harvard.edu/abs/1998ApJ...500..525S} {500, 525}

\bibitem[\protect\citeauthoryear{{Shi} et~al.,}{{Shi} et~al.}{2017}]{Shi2017}
{Shi} D.~D.,  et~al., 2017, \mn@doi [\apj] {10.3847/1538-4357/aa8327}, \href {https://ui.adsabs.harvard.edu/abs/2017ApJ...846...26S} {846, 26}

\bibitem[\protect\citeauthoryear{{Suess} et~al.,}{{Suess} et~al.}{2022}]{Suess2022}
{Suess} K.~A.,  et~al., 2022, \mn@doi [\apj] {10.3847/1538-4357/ac404a}, \href {https://ui.adsabs.harvard.edu/abs/2022ApJ...926...89S} {926, 89}

\bibitem[\protect\citeauthoryear{{Tempel}, {Kruuse}, {Kipper}, {Tuvikene}, {Sorce}  \& {Stoica}}{{Tempel} et~al.}{2018}]{Tempel2018}
{Tempel} E.,  {Kruuse} M.,  {Kipper} R.,  {Tuvikene} T.,  {Sorce} J.~G.,   {Stoica} R.~S.,  2018, \mn@doi [\aap] {10.1051/0004-6361/201833217}, \href {https://ui.adsabs.harvard.edu/abs/2018A&A...618A..81T} {618, A81}

\bibitem[\protect\citeauthoryear{{Trayford}, {Theuns}, {Bower}, {Crain}, {Lagos}, {Schaller}  \& {Schaye}}{{Trayford} et~al.}{2016}]{Trayford2016}
{Trayford} J.~W.,  {Theuns} T.,  {Bower} R.~G.,  {Crain} R.~A.,  {Lagos} C. d.~P.,  {Schaller} M.,   {Schaye} J.,  2016, \mn@doi [\mnras] {10.1093/mnras/stw1230}, \href {https://ui.adsabs.harvard.edu/abs/2016MNRAS.460.3925T} {460, 3925}

\bibitem[\protect\citeauthoryear{{Tremmel}, {Wright}, {Brooks}, {Munshi}, {Nagai}  \& {Quinn}}{{Tremmel} et~al.}{2020}]{Tremmel2020}
{Tremmel} M.,  {Wright} A.~C.,  {Brooks} A.~M.,  {Munshi} F.,  {Nagai} D.,   {Quinn} T.~R.,  2020, \mn@doi [\mnras] {10.1093/mnras/staa2015}, \href {https://ui.adsabs.harvard.edu/abs/2020MNRAS.497.2786T} {497, 2786}

\bibitem[\protect\citeauthoryear{{Tully}}{{Tully}}{2015}]{Tully2015}
{Tully} R.~B.,  2015, \mn@doi [\aj] {10.1088/0004-6256/149/5/171}, \href {https://ui.adsabs.harvard.edu/abs/2015AJ....149..171T} {149, 171}

\bibitem[\protect\citeauthoryear{{Tully}, {Shaya}, {Karachentsev}, {Courtois}, {Kocevski}, {Rizzi}  \& {Peel}}{{Tully} et~al.}{2008}]{Tully2008}
{Tully} R.~B.,  {Shaya} E.~J.,  {Karachentsev} I.~D.,  {Courtois} H.~M.,  {Kocevski} D.~D.,  {Rizzi} L.,   {Peel} A.,  2008, \mn@doi [\apj] {10.1086/527428}, \href {https://ui.adsabs.harvard.edu/abs/2008ApJ...676..184T} {676, 184}

\bibitem[\protect\citeauthoryear{{Wetzel}, {Tinker}, {Conroy}  \& {van den Bosch}}{{Wetzel} et~al.}{2014}]{Wetzel2014}
{Wetzel} A.~R.,  {Tinker} J.~L.,  {Conroy} C.,   {van den Bosch} F.~C.,  2014, \mn@doi [\mnras] {10.1093/mnras/stu122}, \href {https://ui.adsabs.harvard.edu/abs/2014MNRAS.439.2687W} {439, 2687}

\bibitem[\protect\citeauthoryear{{Wild} et~al.,}{{Wild} et~al.}{2014}]{Wild2014}
{Wild} V.,  et~al., 2014, \mn@doi [\mnras] {10.1093/mnras/stu212}, \href {https://ui.adsabs.harvard.edu/abs/2014MNRAS.440.1880W} {440, 1880}

\bibitem[\protect\citeauthoryear{{Wild} et~al.,}{{Wild} et~al.}{2020}]{Wild2020}
{Wild} V.,  et~al., 2020, \mn@doi [\mnras] {10.1093/mnras/staa674}, \href {https://ui.adsabs.harvard.edu/abs/2020MNRAS.494..529W} {494, 529}

\bibitem[\protect\citeauthoryear{{Wong} et~al.,}{{Wong} et~al.}{2012}]{Wong2012}
{Wong} O.~I.,  et~al., 2012, \mn@doi [\mnras] {10.1111/j.1365-2966.2011.20159.x}, \href {https://ui.adsabs.harvard.edu/abs/2012MNRAS.420.1684W} {420, 1684}

\bibitem[\protect\citeauthoryear{{Worthey}, {Faber}, {Gonzalez}  \& {Burstein}}{{Worthey} et~al.}{1994}]{Worthey1994}
{Worthey} G.,  {Faber} S.~M.,  {Gonzalez} J.~J.,   {Burstein} D.,  1994, \mn@doi [\apjs] {10.1086/192087}, \href {https://ui.adsabs.harvard.edu/abs/1994ApJS...94..687W} {94, 687}

\bibitem[\protect\citeauthoryear{{Wright} et~al.,}{{Wright} et~al.}{2010}]{Wright2010}
{Wright} E.~L.,  et~al., 2010, \mn@doi [\aj] {10.1088/0004-6256/140/6/1868}, \href {https://ui.adsabs.harvard.edu/abs/2010AJ....140.1868W} {140, 1868}

\bibitem[\protect\citeauthoryear{{Wright}, {Tremmel}, {Brooks}, {Munshi}, {Nagai}, {Sharma}  \& {Quinn}}{{Wright} et~al.}{2021}]{Wright2021}
{Wright} A.~C.,  {Tremmel} M.,  {Brooks} A.~M.,  {Munshi} F.,  {Nagai} D.,  {Sharma} R.~S.,   {Quinn} T.~R.,  2021, \mn@doi [\mnras] {10.1093/mnras/stab081}, \href {https://ui.adsabs.harvard.edu/abs/2021MNRAS.502.5370W} {502, 5370}

\bibitem[\protect\citeauthoryear{{Yagi}, {Koda}, {Komiyama}  \& {Yamanoi}}{{Yagi} et~al.}{2016}]{Yagi2016}
{Yagi} M.,  {Koda} J.,  {Komiyama} Y.,   {Yamanoi} H.,  2016, \mn@doi [\apjs] {10.3847/0067-0049/225/1/11}, \href {https://ui.adsabs.harvard.edu/abs/2016ApJS..225...11Y} {225, 11}

\bibitem[\protect\citeauthoryear{{Yan}, {Newman}, {Faber}, {Konidaris}, {Koo}  \& {Davis}}{{Yan} et~al.}{2006}]{Yan2006}
{Yan} R.,  {Newman} J.~A.,  {Faber} S.~M.,  {Konidaris} N.,  {Koo} D.,   {Davis} M.,  2006, \mn@doi [\apj] {10.1086/505629}, \href {https://ui.adsabs.harvard.edu/abs/2006ApJ...648..281Y} {648, 281}

\bibitem[\protect\citeauthoryear{{Yan} et~al.,}{{Yan} et~al.}{2009}]{Yan2009}
{Yan} R.,  et~al., 2009, \mn@doi [\mnras] {10.1111/j.1365-2966.2009.15192.x}, \href {https://ui.adsabs.harvard.edu/abs/2009MNRAS.398..735Y} {398, 735}

\bibitem[\protect\citeauthoryear{{Yang}, {Zabludoff}, {Zaritsky}, {Lauer}  \& {Mihos}}{{Yang} et~al.}{2004}]{Yang2004}
{Yang} Y.,  {Zabludoff} A.~I.,  {Zaritsky} D.,  {Lauer} T.~R.,   {Mihos} J.~C.,  2004, \mn@doi [\apj] {10.1086/383259}, \href {https://ui.adsabs.harvard.edu/abs/2004ApJ...607..258Y} {607, 258}

\bibitem[\protect\citeauthoryear{{Yesuf}, {Faber}, {Trump}, {Koo}, {Fang}, {Liu}, {Wild}  \& {Hayward}}{{Yesuf} et~al.}{2014}]{Yesuf2014}
{Yesuf} H.~M.,  {Faber} S.~M.,  {Trump} J.~R.,  {Koo} D.~C.,  {Fang} J.~J.,  {Liu} F.~S.,  {Wild} V.,   {Hayward} C.~C.,  2014, \mn@doi [\apj] {10.1088/0004-637X/792/2/84}, \href {https://ui.adsabs.harvard.edu/abs/2014ApJ...792...84Y} {792, 84}

\bibitem[\protect\citeauthoryear{{Yoon}, {Peterson}, {Kurucz}  \& {Zagarello}}{{Yoon} et~al.}{2010}]{Yoon2010}
{Yoon} J.,  {Peterson} D.~M.,  {Kurucz} R.~L.,   {Zagarello} R.~J.,  2010, \mn@doi [\apj] {10.1088/0004-637X/708/1/71}, \href {https://ui.adsabs.harvard.edu/abs/2010ApJ...708...71Y} {708, 71}

\bibitem[\protect\citeauthoryear{{York} et~al.,}{{York} et~al.}{2000}]{York2000}
{York} D.~G.,  et~al., 2000, \mn@doi [\aj] {10.1086/301513}, \href {https://ui.adsabs.harvard.edu/abs/2000AJ....120.1579Y} {120, 1579}

\bibitem[\protect\citeauthoryear{{Zabludoff}, {Zaritsky}, {Lin}, {Tucker}, {Hashimoto}, {Shectman}, {Oemler}  \& {Kirshner}}{{Zabludoff} et~al.}{1996}]{Zabludoff1996}
{Zabludoff} A.~I.,  {Zaritsky} D.,  {Lin} H.,  {Tucker} D.,  {Hashimoto} Y.,  {Shectman} S.~A.,  {Oemler} A.,   {Kirshner} R.~P.,  1996, \mn@doi [\apj] {10.1086/177495}, \href {https://ui.adsabs.harvard.edu/abs/1996ApJ...466..104Z} {466, 104}

\bibitem[\protect\citeauthoryear{{Zaritsky} et~al.,}{{Zaritsky} et~al.}{2019}]{zaritsky2019}
{Zaritsky} D.,  et~al., 2019, \mn@doi [\apjs] {10.3847/1538-4365/aaefe9}, \href {https://ui.adsabs.harvard.edu/abs/2019ApJS..240....1Z} {240, 1}

\bibitem[\protect\citeauthoryear{{Zaritsky}, {Donnerstein}, {Karunakaran}, {Barbosa}, {Dey}, {Kadowaki}, {Spekkens}  \& {Zhang}}{{Zaritsky} et~al.}{2021}]{zaritsky2021}
{Zaritsky} D.,  {Donnerstein} R.,  {Karunakaran} A.,  {Barbosa} C.~E.,  {Dey} A.,  {Kadowaki} J.,  {Spekkens} K.,   {Zhang} H.,  2021, \mn@doi [\apjs] {10.3847/1538-4365/ac2607}, \href {https://ui.adsabs.harvard.edu/abs/2021ApJS..257...60Z} {257, 60}

\bibitem[\protect\citeauthoryear{{Zaritsky}, {Donnerstein}, {Karunakaran}, {Barbosa}, {Dey}, {Kadowaki}, {Spekkens}  \& {Zhang}}{{Zaritsky} et~al.}{2022}]{zaritsky2022}
{Zaritsky} D.,  {Donnerstein} R.,  {Karunakaran} A.,  {Barbosa} C.~E.,  {Dey} A.,  {Kadowaki} J.,  {Spekkens} K.,   {Zhang} H.,  2022, \mn@doi [\apjs] {10.3847/1538-4365/ac6ceb}, \href {https://ui.adsabs.harvard.edu/abs/2022ApJS..261...11Z} {261, 11}

\bibitem[\protect\citeauthoryear{{Zaritsky}, {Donnerstein}, {Dey}, {Karunakaran}, {Kadowaki}, {Khim}, {Spekkens}  \& {Zhang}}{{Zaritsky} et~al.}{2023}]{zaritsky2023}
{Zaritsky} D.,  {Donnerstein} R.,  {Dey} A.,  {Karunakaran} A.,  {Kadowaki} J.,  {Khim} D.~J.,  {Spekkens} K.,   {Zhang} H.,  2023, \mn@doi [\apjs] {10.3847/1538-4365/acdd71}, \href {https://ui.adsabs.harvard.edu/abs/2023ApJS..267...27Z} {267, 27}

\bibitem[\protect\citeauthoryear{{Zhang}, {Li}, {Leja}, {Whitaker}, {Nersesian}, {Bezanson}  \& {van der Wel}}{{Zhang} et~al.}{2023}]{Zhang2023}
{Zhang} J.,  {Li} Y.,  {Leja} J.,  {Whitaker} K.~E.,  {Nersesian} A.,  {Bezanson} R.,   {van der Wel} A.,  2023, \mn@doi [\apj] {10.3847/1538-4357/acd84a}, \href {https://ui.adsabs.harvard.edu/abs/2023ApJ...952....6Z} {952, 6}

\bibitem[\protect\citeauthoryear{{Zou} et~al.,}{{Zou} et~al.}{2017}]{Zhou2017}
{Zou} H.,  et~al., 2017, \mn@doi [\pasp] {10.1088/1538-3873/aa65ba}, \href {https://ui.adsabs.harvard.edu/abs/2017PASP..129f4101Z} {129, 064101}

\bibitem[\protect\citeauthoryear{{de Blok} \& {McGaugh}}{{de Blok} \& {McGaugh}}{1997}]{deBlok1997}
{de Blok} W.~J.~G.,  {McGaugh} S.~S.,  1997, \mn@doi [\mnras] {10.1093/mnras/290.3.533}, \href {https://ui.adsabs.harvard.edu/abs/1997MNRAS.290..533D} {290, 533}

\bibitem[\protect\citeauthoryear{{van Dokkum}, {Abraham}, {Merritt}, {Zhang}, {Geha}  \& {Conroy}}{{van Dokkum} et~al.}{2015a}]{vanDokkum2015a}
{van Dokkum} P.~G.,  {Abraham} R.,  {Merritt} A.,  {Zhang} J.,  {Geha} M.,   {Conroy} C.,  2015a, \mn@doi [\apjl] {10.1088/2041-8205/798/2/L45}, \href {https://ui.adsabs.harvard.edu/abs/2015ApJ...798L..45V} {798, L45}

\bibitem[\protect\citeauthoryear{{van Dokkum} et~al.,}{{van Dokkum} et~al.}{2015b}]{vanDokkum2015b}
{van Dokkum} P.~G.,  et~al., 2015b, \mn@doi [\apjl] {10.1088/2041-8205/804/1/L26}, \href {https://ui.adsabs.harvard.edu/abs/2015ApJ...804L..26V} {804, L26}

\bibitem[\protect\citeauthoryear{{van Dokkum} et~al.,}{{van Dokkum} et~al.}{2016}]{vanDokkum2016}
{van Dokkum} P.,  et~al., 2016, \mn@doi [\apjl] {10.3847/2041-8205/828/1/L6}, \href {https://ui.adsabs.harvard.edu/abs/2016ApJ...828L...6V} {828, L6}

\bibitem[\protect\citeauthoryear{{van der Burg}, {Muzzin}  \& {Hoekstra}}{{van der Burg} et~al.}{2016}]{vanderBurg2016}
{van der Burg} R. F.~J.,  {Muzzin} A.,   {Hoekstra} H.,  2016, \mn@doi [\aap] {10.1051/0004-6361/201628222}, \href {https://ui.adsabs.harvard.edu/abs/2016A&A...590A..20V} {590, A20}

\bibitem[\protect\citeauthoryear{{van der Burg} et~al.,}{{van der Burg} et~al.}{2017}]{vanderBurg2017}
{van der Burg} R. F.~J.,  et~al., 2017, \mn@doi [\aap] {10.1051/0004-6361/201731335}, \href {https://ui.adsabs.harvard.edu/abs/2017A&A...607A..79V} {607, A79}

\makeatother
\end{thebibliography}


\label{lastpage}
\end{document}